\documentclass[aps,prx,onecolumn,showpacs,amsfonts,amssymb,floatfix]{revtex4-2}
\usepackage[T1]{fontenc}
\usepackage[utf8]{inputenc}
\usepackage{float}
\usepackage{amssymb}
\usepackage{amsmath}
\usepackage{enumerate}
\usepackage{orcidlink}
\usepackage{hyperref}
\usepackage{cleveref}
\usepackage{mathtools}
\usepackage{color}
\usepackage{etoolbox}
\usepackage{graphicx}
\usepackage{dcolumn}
\usepackage{bbm}
\usepackage{accents}
\newlength{\dhatheight}
\newcommand{\doublehat}[1]{%
    \settoheight{\dhatheight}{\ensuremath{\hat{#1}}}%
    \addtolength{\dhatheight}{-0.35ex}%
    \hat{\vphantom{\rule{1pt}{\dhatheight}}%
    \smash{\hat{#1}}}}

\makeatletter

\newcommand*{\ro}{\mathbf{r}_1}
\newcommand*{\Ctwo}{KSZI20,KI20,KI21}
\newcommand*{\flocking}{PRL75_VCBCS95,PRL75_TT95,PRE58_TT98,PDB08,PSB10,SCT15,PRL119_LL17,MLDP18,PRR1_LPL19,PRL123_MGC19,KI20,KI21}
\newcommand*{\nematicflocking}{PRE74_PDB06,PRL101_BM08,PRL104_GPBC10,EPJ225_Peruani16,ARCMP_BGHP20,EPL143_KL23}
\newcommand*{\percolation}{QZ07,MM12,KBK23}
\newcommand*{\bands}{SCT15,MLDP18,KI20}
\newcommand*{\bbgky}{BG_1946,Bogoliubov_1946,Kirkwood_1946,Kirkwood_1947}
\newcommand*{\ursell}{Ursell_1927}
\newcommand*{\correlations}{KSZI20,Kuersten19}
\newcommand*{\ringkin}{PRE91_CI15, KI21}
\newcommand*{\kinmf}{PRE86_CWI12}
\newcommand*{\Peclet}{P\'eclet }

\newcount\hour \newcount\minute
\hour=\time \divide \hour by 60 \minute=\time \count99=\hour
\multiply \count99 by -60 \advance\minute by \count99 \hour=\time
\divide \hour by 60
\date{February 28, revised: March 8, 2024} 

\makeatother

\begin{document}

\title{Universal Scaling of Clustering Instability for Interacting Active Brownian Particles}

\author{Rüdiger Kürsten\, \orcidlink{0000-0001-5830-0442}}
\affiliation{Departament de Física de la Matèria Condensada, Universitat de Barcelona, Martí i Franquès 1, 08028 Barcelona, Spain}
\affiliation{UBICS University of Barcelona Institute of Complex Systems - C. Martí i Franquès 1, E08028 Barcelona, Spain}
\affiliation{Institut für Physik, Universität Greifswald, Felix-Hausdorff-Str. 6, 17489 Greifswald, Germany}
\email{kursten@ub.edu}

\begin{abstract}
Clustering is one of the mayor collective phenomena observed in active matter.
We study the overdamped motion of interacting active Brownian particles in two dimensions.
An instability in the pair correlation function causes the onset of clustering.
This clustering mechanism depends mainly on the self-propulsion properties of the active particles and details of the interactions do not effect the scaling of the clustering instability.
Theoretical predictions from repeated ring-kinetic theory are confirmed by agent-based simulations.
\end{abstract}
\maketitle

\section{Introduction}

Active matter is characterized by the interplay between directed motion and dissipation of active units.
Many natural or artificial objects are active. 
Examples are humans, robots, animals, bacteria or active colloids.
Collections of many active units exhibit fascinating collective phenomena.
Due to the inherent non-equilibrium nature of active matter, its collective phases can be significantly different from equilibrium phases.

Typical collective phenomena observed in large collections of self-propelled particles are flocking, clustering and motility induced phase separation (MIPS).

By flocking we mean the self-organization into a globally polarized state such as observed in flocks of birds \cite{\flocking}.
Typical flocking models are based on microscopic interactions that favor alignment between nearby particles.
However, flocking is also observed in models without explicit alignment interactions due to e.g. purely repulsive  \cite{PRL124_CMP20,PRE106_CL22} or attractive \cite{PRL130_CL23} interactions, memory effects \cite{PRE107_KK23, EPL143_KK23}, purely anti-aligning interactions \cite{ARX_KMI23}, by turning away \cite{ARX_DCZYZA24}, or due to entropy maximization \cite{PRL130_DT23}.
By flocking in a wider sense we might refer to the preferred motion into two \cite{\nematicflocking} (nematic order), or more directions \cite{PRE90_RLI14,PRL119_KSI17}.

By clustering we denote the appearance of spatially inhomogeneous particle distributions such that there are many local accumulations of particles in different places. Clustering is observed, for example, in groups of animals \cite{AB67_HCGK04,CB10_ACDGMPPRSSV14}, bacteria \cite{PRL108_PSJSDB12} or active colloids \cite{PRL108_TCPYB12,PRL110_BBKLBS13}.
In models, clustering is found for both, excluded volume interactions \cite{PRL131_CCDGLS23,PRR3_HKF21,PRE74_PDB06,JNCS407_BSL15,PRL110_RHB13,ATS3_MYN20,PRE99_ANVP19,PRL112_ZS14} and models with pure alignment interactions \cite{ZHR22,EPJST224_GRBS15,PRL92_HA04,EPJST191_PSB10,NJP15_PB13}.
Clustering is often described by means of heuristic kinetic clustering theories \cite{PRR3_HKF21,PRE74_PDB06,EPJST191_PSB10,NJP15_PB13} where scattering amplitudes have to be estimated either from principle physical mechanisms or from simulation data.
In many cases, power law distributed cluster sizes are found and there can be a transition towards the presence of a single huge cluster.
A state with a giant dense cluster surround by a disordered gas is known as MIPS \cite{ARCMP6_CT15,OSTZ23}.

In this study we intend to understand the onset of clustering from first principles.
In general, clustering is a correlation effect that can hardly be understood at the level of the distribution of a single particle.
In dense systems, as for example, dense active colloidal systems, clusters are composed of many particles and clearly correlations between more than two particles are essential in order to describe the cluster.
Such complicated clusters that involve interactions between many particles at the same time are hardly rigorously feasible and the phenomenological kinetic cluster theories mentioned above are probably the most convenient choice to describe such clustering states.
On the other hand, experiments like in \cite{PRL108_TCPYB12} show that clustering does occur already at relatively small packing fractions of about $3\%$.
Furthermore, at least at not too large densities, it is reasonable to assume that clusters form from disorder, first, by collisions of pairs of particles.
In that way, in a first step, large pair correlations are formed and larger clusters accumulate in a second step via more complicated collisions between more than two particles.

In this paper we develop an understanding of the onset of clustering from first principles and thus complement previous phenomenological works on the cluster size distribution.
We study active Brownian particles (ABPs) \cite{EPJST202_RBELS12} in two dimensions as a prototype model for self-propelled particles.
For simplicity, we consider the limit of overdamped ABPs with high activity.
We study clustering under the impact of alignment interactions between neighboring particles.
We consider two standard definitions of neighborhoods: metric neighborhoods that are all particles in a given distance and topological neighborhoods that consist of a given number of closest neighbors.
In both cases, the considered model depends on three dimensionless parameters: \Peclet number $Pe$ that controls the activity, coupling strength $\Gamma$ and density $M$.

We develop a repeated ring-kinetic theory \cite{\ringkin}, that is a set of two nonlinear equations analogous to the nonlinear Fokker-Planck equation of mean field theories.
Here however, we take pair correlations fully into account.
We identify a linear instability in the pair correlation function at the onset of clustering.
Remarkably, this instability is essentially determined by the ABP-operator (describing the action of self-propulsion and orientational diffusion) only and does not depend on the details of the interactions.
In the limit of large \Peclet numbers, the onset of clustering depends on the ratio $Pe/\Gamma^{1.6}$, that is close to the exponent of $1.5$ that was measured in simulations of a similar, topological model in Ref. \cite{ZHR22}.
For small \Peclet numbers we find a different, long wavelength instability of the pair correlation function that dominates in this regime and creates large scale patterns.
Theoretical predictions are confirmed in agent-based simulations.

The paper is organized as follows.
In Sec. \ref{sec:model} we define the models under consideration with dimensional parameters.
We then perform a rescaling of space and time to arrive at a reduced number of only three dimensionless parameters, compatible for both considered models (metric and topological neighborhood definition).
In Sec. \ref{sec:simulations} we study clustering in both models in agent-based simulations of $N=10^4$ particles.
We consider the mean cluster size and a local integral of the pair correlation function as order parameters.
We systematically study the onset of clustering in dependence on \Peclet number and interaction strength.
In Sec. \ref{sec:ringkin} we derive the repeated ring-kinetic theory for the topological model from first principles, going beyond mean field theories of similar models \cite{\kinmf}. The repeated ring-kinetic theory for the metric model has been derived in Ref. \cite{KI21}, we give the corresponding equations here, within the notation of this paper.
In Sec. \ref{sec:instab} we identify the instabilities of the pair correlation functions that lead to clustering or large scale patterns.
In Sec. \ref{sec:discussion} we summarize the findings of this paper and discuss its implications.

\section{Model\label{sec:model}}

We study $N$ point particles that move at constant speed $v$ in the two-dimensional domain $(0, \tilde{L}_x]\times (0,\tilde{L}_y]$ under periodic boundary conditions. Thus the global density is given by $\tilde{\rho}=N/(\tilde{L}_x\tilde{L}_y)$.
We consider the following equation of motion
\begin{align}
	\frac{d}{d \tilde{t}}\tilde{x}_i(\tilde{t}) = v \cos \tilde{\phi}_i(\tilde{t}),\,\,\, 
	\frac{d}{d \tilde{t}}\tilde{y}_i(\tilde{t}) = v \sin \tilde{\phi}_i(\tilde{t}),
	\label{eq:self_propulsion_xy}
\end{align}
where $\mathbf{\tilde{r}}_i=(\tilde{x}_i, \tilde{y}_i)$ denotes the position of particle $i$ and the angle $\tilde{\phi}_i\in[-\pi, \pi]$ its direction of self-propulsion.
The orientations $\tilde{\phi}_i$ undergo the following Langevin-dynamics
\begin{align}
	\frac{d}{d \tilde{t}}\tilde{\phi}_i(\tilde{t})=\tilde{\Gamma}w(|\Omega_i|) \sum_{j\in \Omega_i}K[\tilde{\phi}_j(\tilde{t})-\tilde{\phi}_i(\tilde{t})] + \sigma\cdot \tilde{\xi}_i(\tilde{t}),
	\label{eq:langevin_orientation}
\end{align}
where $\tilde{\Gamma} \ge 0$ is the alignment strength and the $\tilde{\xi}_i$ denote independent Gaussian delta-correlated noise $\langle \tilde{\xi}_i(\tilde{t}) \tilde{\xi}_j(\tilde{s})\rangle=\delta_{ij}\delta(\tilde{t}-\tilde{s})$.
The function $K$ defines the actual interaction of orientations of neighboring particles.
In this paper we consider only the case $K(\Delta \phi):=\sin(\Delta \phi)$, however, we develop the theory for arbitrary functions $K$.
By $\Omega_i$ we denote the set of indexes of particles that we consider 'neighbors' of particle $i$. 
The weight function $w(|\Omega_i|)$ depends on the cardinality of the set $\Omega_i$ that is the number of neighbors of particle $i$. 
In this paper we consider either the weight function $w(n)\equiv 1$ or $w(n)=1/n$ but we formulate the theory for arbitrary weight functions $w$.
The neighborhood might be defined in various ways and we consider two distinct types of neighborhood definitions.

In what is known as \textit{topological} or metric-free neighborhood, $\Omega_i$ embraces $i$ itself, and the $M$ closest particles to particle $i$ (different from particle $i$). 
In the simplest case of $M=1$, $\Omega_i$ contains only $i$ and the index of the particle closest to particle $i$.
If this particle has index $j$ this means $\Omega_i=\{i, j\}$.
Drawing the Voronoi diagram of particles $\{1, 2, \dots, i-1, i+1, \dots, N\}$, particle $i$ lies in the Voronoi cell of particle $j$. 
For $M=2$, the second closest neighbor of particle $i$ can be found by looking at the Voronoi diagram of particles $\{1, 2, \dots, N\}/\{i,j\}$. 
If particle $i$ lies in the Voronoi cell of particle $l$, then $\Omega_i=\{i, j, l\}$.
One proceeds analogously for $M>2$.
Note that interactions defined in this way are not reciprocal.
If particle $j\in \Omega_i$ such that $\tilde{\phi}_i$ is affected by $\tilde{\phi}_j$ it is not necessarily true that also $i\in \Omega_j$ that means that $\tilde{\phi}_j$ is also affected by $\tilde{\phi}_i$. 
For the topological neighborhood definition, the weight function $w(|\Omega_i|)=w(1+M)$ is constant once the parameter $M$ is specified.
Without loss of generality, we consider $w\equiv 1$ in this case.

In what is known as \textit{metric} neighborhood, $\Omega_i$ embraces the indexes of all particles that have a distance less or equal to $\tilde{R}$ to particle $i$ (including $i$ itself). 
Thus $\Omega_i=\{j: \sqrt{(\tilde{x}_i-\tilde{x}_j)^2+(\tilde{y}_i-\tilde{y}_j)^2}\le \tilde{R}\}$.
For metric neighborhoods, mainly the weight functions $w(n)\equiv 1$ (additive interactions) and $w(n):=1/n$ (non-additive interaction) have been considered in the literature.
It is known that the two models behave qualitatively different \cite{CSP21,KI21} and that the model with non-additive interactions is more similar to the seminal Vicsek model.
Therefore, we consider the non-additive interactions here, however we keep an arbitrary weight function $w$ in the theory.

Note that in the definitions of both, topological and metric neighborhoods, we included the focal particle itself in the neighborhood set, that is $i\in \Omega_i$.
However, this choice is physically not relevant because the corresponding interaction term does not contribute in the equation of motion because $K(\phi_i-\phi_i)=K(0)=0$ for all considered interaction functions $K$.
Thus, an alternative definition of the neighborhood, excluding the focal particle, is mathematical equivalent if also the weight function $w$ is defined correspondingly (shifted by one).

We might consider a rescaling of time and the noise processes $\tilde{\xi}_i$ according to
\begin{align}
	\phi_i(t)=\tilde{\phi}_i(\tilde{t}), \,\,\, t=\sigma^2\tilde{t}, \,\,\, \xi_i(t)= \tilde{\xi}_i(\tilde{t}) /\sigma,
	\label{eq:rescaling_time}
\end{align}
such that the new noise processes are delta-correlated with unit variance in the new time $\langle \xi_i(t)\xi_j(s)\rangle= \delta_{ij}\delta(t-s)$. 
Furthermore, we rescale space for the topological neighborhood definition according to
\begin{align}
	x_i(t)=\tilde{x}_i(\tilde{t})\sqrt{\tilde{\rho}}, \, \, \, y_i(t)=\tilde{y}_i(\tilde{t})\sqrt{\tilde{\rho}}
	\label{eq:rescaling_space_topological}
\end{align}
and for the metric neighborhood definition according to
\begin{align}
	x_i(t)=\tilde{x}_i(\tilde{t})/\tilde{R}, \, \, \, y_i(t)=\tilde{y}_i(\tilde{t})/\tilde{R}.
	\label{eq:rescaling_space_metric}
\end{align}
With the above transformation the equations of motion become
\begin{align}
	\frac{d}{dt}x_i(t)=Pe \cos\phi_i(t), \,\,\, \frac{d}{dt}y_i(t)= Pe \sin \phi(t), \,\,\, \frac{d}{dt} \phi_i(t)=\Gamma w(|\Omega_i|) \sum_{j\in \Omega_i} K[\phi_j(t)-\phi_i(t)]+ \xi_i(t),
	\label{eq:langevin}
\end{align}
with the three dimensionless microscopic parameters \Peclet number $Pe$, coupling $\Gamma$ and (mean) neighbor number $M$.
The rescaled spatial domain is denoted by $(x_i,y_i)\in (0,L_x]\times (0,L_y]$. 

For the topological model, the dimensionless parameters are given in terms of the unscaled parameters according to $Pe=v\sqrt{\tilde{\rho}}/\sigma^2$ and $\Gamma=\tilde{\Gamma}/\sigma^2$ and the neighbor number $M$ is a fixed integer parameter.
The rescaled domain boundaries are $L_x:=\sqrt{N}\sqrt{\tilde{L}_x}/\sqrt{\tilde{L}_y}$ and $L_y:=\sqrt{N}\sqrt{\tilde{L}_y}/\sqrt{\tilde{L}_x}$.
Thus, the average global density in the rescaled variables is equal to one, $\rho=N/(L_xL_y)=1$.

For the metric model, the dimensionless parameters are given in terms of the unscaled parameters according to $Pe=v/(\tilde{R}\sigma^2)$, $\Gamma=\tilde{\Gamma}/\sigma^2$ and the mean neighbor number $M=\pi R^2 \rho=\pi \tilde{R}^2 \tilde{\rho}$, where $R=1$, $L_x=\tilde{L}_x/\tilde{R}=\sqrt{N}\sqrt{\pi/M}\sqrt{\tilde{L}_x}/\sqrt{\tilde{L}_y}$, $L_y=\tilde{L}_y/\tilde{R}=\sqrt{N}\sqrt{\pi/M}\sqrt{\tilde{L}_y}/\sqrt{\tilde{L}_x}$, $\rho=M/\pi=N \tilde{R}^2/(\tilde{L}_x\tilde{L}_y)$ and the neighborhoods are given by the unit circle $\Omega_i=\{j: \sqrt{(x_i-x_j)^2+(y_i-y_j)^2}\le 1 \}$.
Note that in case of the metric model, the mean neighbor number, parameter $M$, can take each positive real number as a value, not just integers as in the topological model.

In summary, the reduced equation of motion, \cref{eq:langevin} depends on the three dimensionless microscopic (intensive) parameters of the dynamics, \Peclet number $Pe$, coupling strength $\Gamma$ and neighbor number $M$. Furthermore, the system is characterized by the aspect ratio of the considered domain $L_x/Ly$ and the system size given by the particle number $N$.
The rescaled equation of motion, \cref{eq:langevin}, is mathematical equivalent to the unscaled equation of motion, \cref{eq:self_propulsion_xy,eq:langevin_orientation} that depends on more, dimensional, parameters.
Throughout the rest of the paper we are employing only the formulation \cref{eq:langevin}.

\section{Clustering in Agent Based Simulations \label{sec:simulations}}

We perform agent-based simulations of \cref{eq:langevin} for $N=10^4$ particles in a quadratic domain with periodic boundary conditions.
We employ the Euler-Maruyama scheme with step size $\Delta t=10^{-4}$ in all cases and perform $10^7$ time steps after which the system is thermalized.
All simulations have been done employing the AAPPP package \cite{aappp} for the actual dynamics and the NetworkX package \cite{networkx} for network analysis.
We consider various values of \Peclet number and coupling $Pe, \Gamma \in \{2^{-2}, 2^{-1}, \dots, 2^{11}\}$.
All simulations are started with spatially homogeneous and isotropic random initial conditions.
We use the harmonic interaction function $K(\Delta \phi):=\sin(\Delta \phi)$.

\begin{figure}[h]
	\begin{center}
		\includegraphics[width=0.19\textwidth]{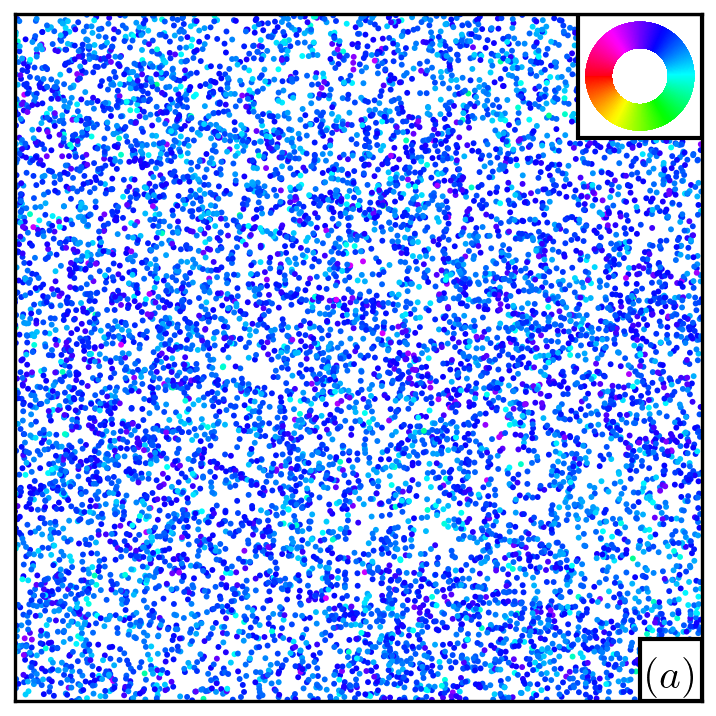}
		\includegraphics[width=0.19\textwidth]{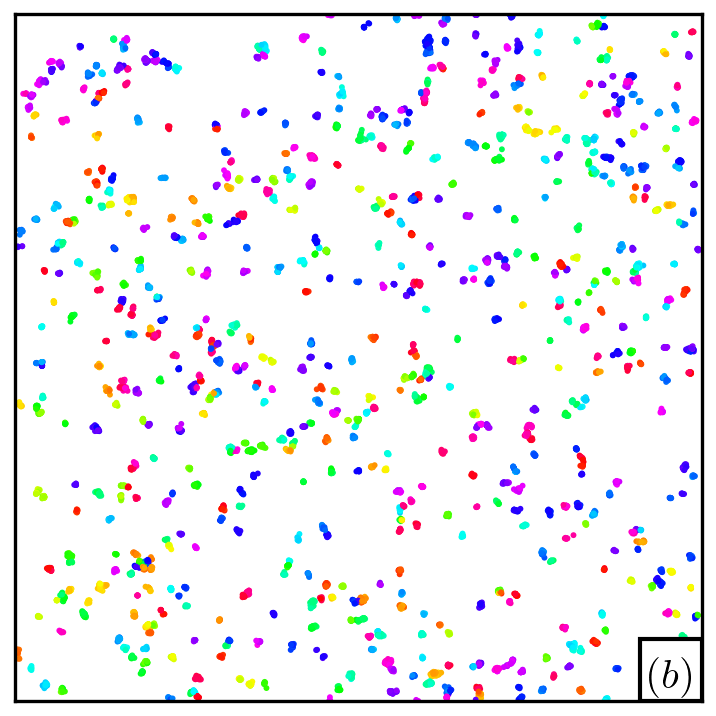}
		\includegraphics[width=0.19\textwidth]{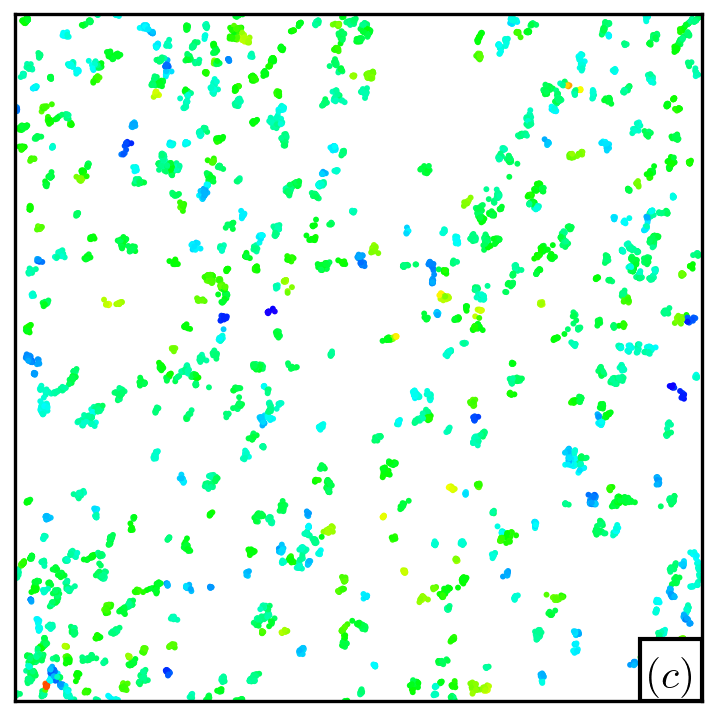}
		\includegraphics[width=0.19\textwidth]{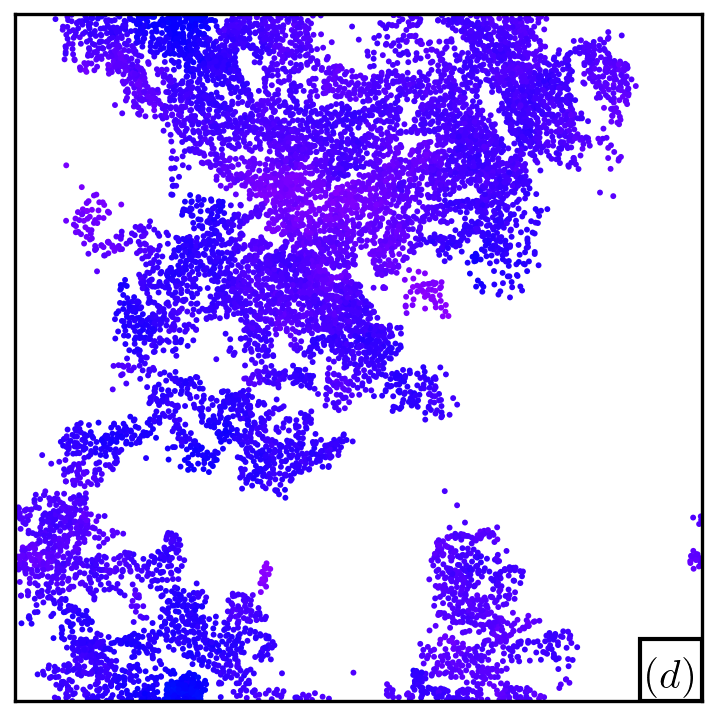}
		\includegraphics[width=0.19\textwidth]{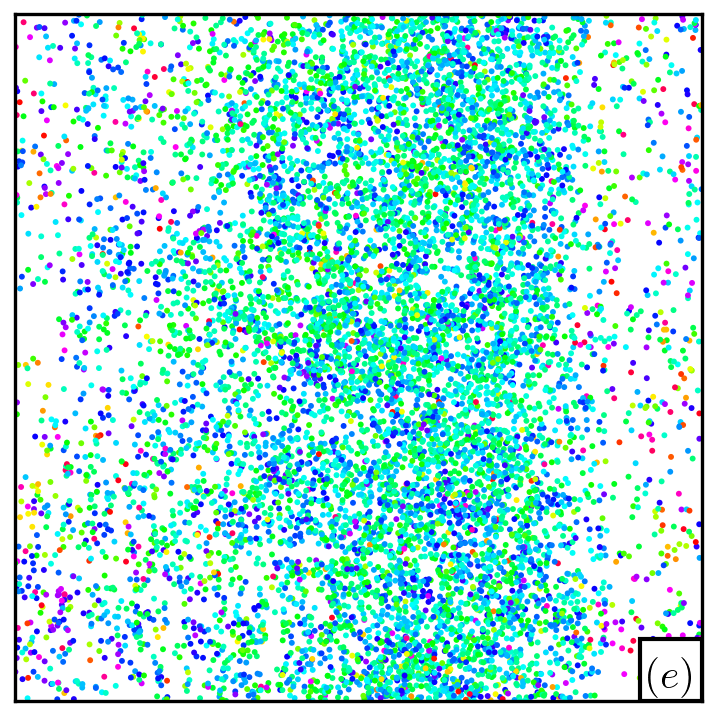}
	\end{center}
	\caption{Snapshots taken after $10^7$ steps with step size $\Delta t=10^{-4}$ for the $(a-d)$ topological neighborhood definition and metric neighborhood definition $(e)$. 
	Particle orientations are encoded in color according to the color wheel shown in $(a)$.
	The displayed phases are $(ii)$ homogeneous polarized phase $[Pe=128, g=1, (a)]$, $(iii)$ isotropic clustering phase $[Pe=2, g=16, (b)]$, $(iv)$ polarized clustering phase $[Pe=16, g=32, (c)]$, $(v)$ polarized percolating phase $[Pe=1, g=256, (d)]$ and $(vi)$ band phase $[Pe=8, g=2, (e)]$.
	The disordered phase $(i)$ is not shown.
	The (expected) number of neighbors is $M=5$ in $(a-d)$ and $M=4.6$ in $(e)$.
	All simulations have been started with random initial conditions, the system size is given by $N=10^4$.}
	\label{fig:phases}
\end{figure}

For the topological neighborhood definition we use, without loss of generality, the weight function $w(n)\equiv 1$.
Because the number of neighbors $n=1+M$ is constant, different weight functions might be absorbed into the coupling constant $\Gamma$.
Quantitatively, our simulations for the topological model reproduce the results of Ref. \cite{ZHR22} where the piecewise linear saw-tooth function was used as interaction function and $N=2500$, $M=3$ was used in simulations.
However, our simulations provide some additional insides.
We show simulation results for $M=5$, but we also performed simulations for $M=1, 2, \dots, 7$ with similar results (not shown).
Depending on \Peclet number and coupling we observe the typical states (phases): $(i)$ disorder (not shown), $(ii)$ homogeneous polarized state, see Fig.\ref{fig:phases} $(a)$, $(iii)$ clustered isotropic state, see Fig.\ref{fig:phases} $(b)$, $(iv)$ clustered polarized state, see Fig.\ref{fig:phases} $(c)$ and $(v)$ a patchy, polarized percolating state, see Fig.\ref{fig:phases} $(d)$.

Within the disordered phase $(i)$, particles are homogeneously distributed in space and their orientations are distributed isotropically.
Within the homogeneous polarized phase $(ii)$, the spatial distribution is still homogeneous, but particles move on average into a preferred direction.
Within the clustered isotropic phase $(iii)$, particles spatially arrange into small groups (clusters) of few, locally aligned particles. 
However, those clusters are oriented into different, random directions, such that there is no global polar order.
Within the clustered polarized phase $(iv)$, those clusters get orientationally ordered such that there is global polar order.
Within the polarized patchy state $(v)$, there is basically one big (or very few) polarized cluster that contains almost all particles.
Although the huge cluster percolates the domain, it does not occupy all space and thus the spatial distribution is not homogeneous.
The phases $(i)-(iv)$ have been observed and discussed in \cite{ZHR22}, phase $(v)$ was not reported therein, it might have been out of the considered parameter range of \cite{ZHR22}.

\begin{figure}[h]
	\begin{center}
		\includegraphics[width=0.49\textwidth]{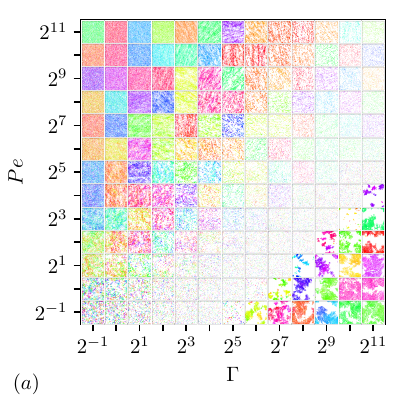}
		\includegraphics[width=0.49\textwidth]{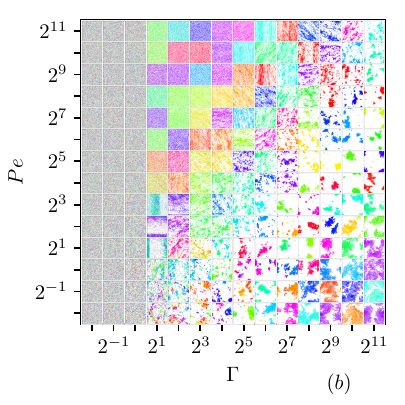}
	\end{center}
	\caption{Overview of snapshots for different values of \Peclet number $Pe$ and coupling $\Gamma$  for $(a)$ topological neighborhoods at $M=5$ and $(b)$ metric neighborhoods at $M=4.6$.
	Colors indicate particle orientations as in the color wheel of Fig.\ref{fig:phases} $(a)$, parameters as in Fig.\ref{fig:phases}.}
	\label{fig:snapshots}
\end{figure}

In Fig. \ref{fig:snapshots} $(a)$ we display snapshots for the topological neighborhood definition with $M=5$, for all considered values of the parameters $Pe$ and $\Gamma$. 

\begin{figure}[h]
	\begin{center}
		\includegraphics[width=0.49\textwidth]{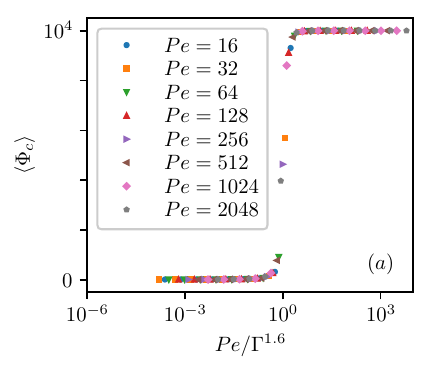}
		\includegraphics[width=0.49\textwidth]{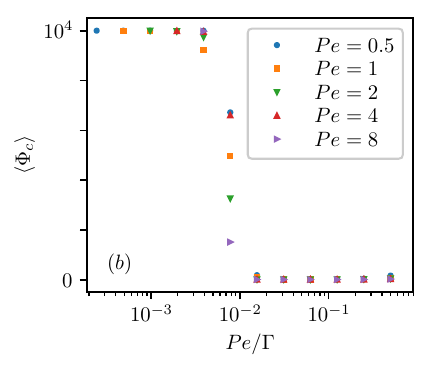}
	\end{center}
	\caption{Collapse of mean cluster size at different \Peclet numbers for the topological neighborhood definition.
	Cluster size order parameter $\langle \Phi_c \rangle$, for different values of $Pe$ and coupling $\Gamma$ for $(a)$ high \Peclet numbers as a function of $Pe/\Gamma^{1.6}$ and for $(b)$ small \Peclet numbers as a function of $Pe/\Gamma$.
For each data point a hundred measurements have been taken each unit of time ($10^4$ steps) for after a thermalization time $t_0=10^3$ ($10^7$ time steps) for one realization. 
	}
	\label{fig:collapse1}
\end{figure}

Flocking in the considered or similar models has been studied in many works, see e.g. \cite{\flocking}.
Here we are focusing on the study of clustering in more detail.
For this purpose we are following the graph-theoretic approach used in Ref. \cite{ZHR22}.
We consider a graph of $N$ nodes, representing the $N$ particles.
We connect nodes $i$ and $j$ if either $i$ is a neighbor of $j$ or if $j$ is a neighbor of $i$, that is if $i\in \Omega_j$ or $j\in \Omega_i$.
Following \cite{ZHR22} we consider the expected cluster size of a randomly chosen node (particle) $\langle \Phi_c\rangle$ as an order parameter.
In phases $(i)$ and $(ii)$ particles are distributed more or less homogeneously, that means the particle positions are close to be distributed independently at random.
Evidently, for randomly distributed particles with topological neighborhoods and $M=5$, the corresponding graph is percolating.
Thus, there is a giant cluster that contains almost all particles and the order parameter $\langle \Phi_c \rangle$ is close to $N$.
In phases $(iii)$ and $(iv)$ particles assemble in small groups and thus form a large number of clusters of about the same size, which is around $M$ that is close to zero in comparison to the particle number $N$.
Thus, depending on $Pe$ and $\Gamma$, there is a transition from $\langle \Phi_c \rangle \approx N$ in phases $(i-ii)$ towards $\langle \Phi_c \rangle \approx 0$ in phases $(iii-iv)$.
It was reported in Ref. \cite{ZHR22} that the value of the order parameter $\langle \Phi_c \rangle$ does not depend on the parameters $Pe$ and $\Gamma$ individually, but only on the fraction $Pe/\Gamma^{\alpha}$, where $\alpha\approx 1.5$ was given in \cite{ZHR22} (with slightly different interactions and for a smaller system size as we use).
In Fig. \ref{fig:collapse1} $(a)$ we see the clustering transition as expected.
We find a collapse of curves for different $Pe$ using the rescaling by $\Gamma^{\alpha}$ with $\alpha=1.6$.
We have theoretical arguments for this value of the exponent for high \Peclet numbers, see Sec. \ref{sec:instab}.
However, the resolution of our data does not allow to securely distinguish between $\alpha=1.5$ and $\alpha=1.6$. 

Similarly to phases $(i-ii)$, in phase $(v)$ the particles are distributed approximately homogeneous and there is a giant cluster in the corresponding neighborhood graph.
Thus, there is another transition between phases $(iii-iv)$ $[\langle \Phi_c \rangle \approx 0]$ to phase $(v)$ $[\langle \Phi_c \rangle \approx N]$ at small \Peclet numbers where we find phase $(v)$ within the considered parameter regime.
Again, we find a collapse of curves for different $Pe$ using the rescaling by $\Gamma^{\alpha}$ but in this case with $\alpha=1.0$, see Fig. \ref{fig:collapse1} $(b)$.
We do gain an understanding also for this exponent, see Sec. \ref{sec:instab}.
We did obtain very similar results for $M=1, \dots, 7$ (not shown), however, for $M=1,2$ the clustering transition does not go along with a percolation transition because even for independent randomly distributed particles there is no giant cluster in this case.

Next, we study the dynamics with metric interactions for $M=4.6$.
The phase diagram as well as the clustering phenomena behave qualitatively equivalent for other values of $M\in [0.2, 12.8]$ (not shown), however, the clustering phenomena do not go along with a percolation transition in all cases.
The density chosen is slightly above the percolation threshold of uncorrelated discs in two dimensions, see e. g. \cite{\percolation}.
In the case of metric neighborhoods, we find in principle the same phases as for the topological model, however, there is an additional phase $(vi)$ of polarized high density bands surrounded by a disordered gas of much lower density, see Fig. \ref{fig:phases} $(e)$.
The presence of such bands in Vicsek-like models, as well as in exactly the model considered here is known for many years, see e.g. \cite{\bands}.
In fact, it is known that the homogeneous polarized state $(ii)$ goes over into the band phase $(vi)$ for larger system sizes.
For the Vicsek model with topological neighborhood it is known that there are also band states for large enough systems, see Ref. \cite{PRL126_MCNSTW21}.
For the topological Langevin model considered here, we do not observe band states for any considered parameter set, thus it remains an open question whether or not band states exist for large enough system sizes.
We display an overview of snapshots for all considered parameters for the metric neighborhood model in Fig. \ref{fig:snapshots} $(b)$.
\begin{figure}[h]
	\begin{center}
		\includegraphics[width=0.49\textwidth]{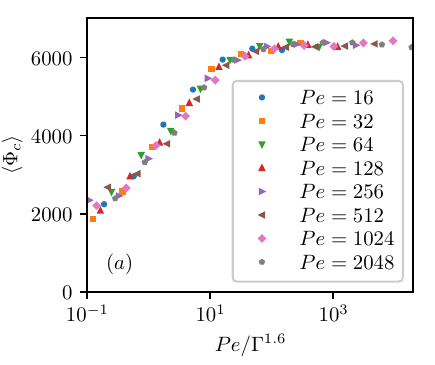}
		\includegraphics[width=0.49\textwidth]{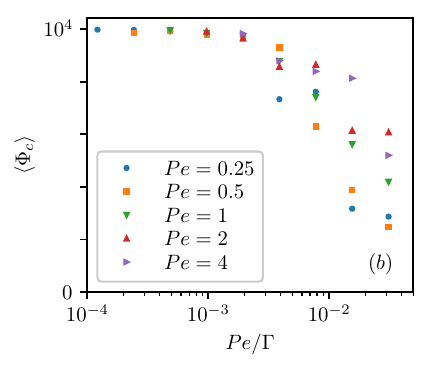}\\
		\includegraphics[width=0.49\textwidth]{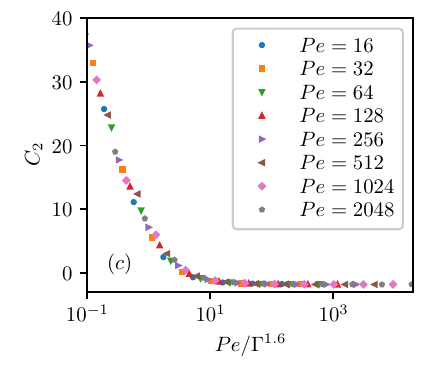}
		\includegraphics[width=0.49\textwidth]{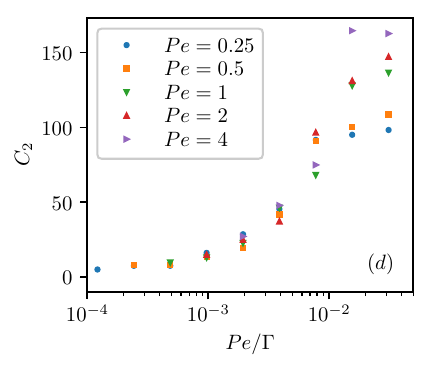}
	\end{center}
	\caption{Collapse of mean cluster size $(a-b)$ and local integral of the radial distribution function, $C_2$, $(c-d)$ at different \Peclet numbers for metric neighborhood definition.
	For high \Peclet numbers, clustering depends on $Pe/\Gamma^{1.6}$ $(a), (c)$ and for small \Peclet numbers it depends on $Pe/\Gamma$ $(b), (d)$.
	For each data point a hundred measurements have been taken each unit of time ($10^4$ steps) for after a thermalization time $t_0=10^3$ ($10^7$ time steps) and averaged over five realizations. }
	\label{fig:collapse2}
\end{figure}

Considering the neighborhood graph, we find find a percolating cluster in phases $(i-ii)$ as for the topological model.
Here however, the percolating cluster does not contain all particles but only a bit more than $6000$ from $N=10000$.
The transition towards clustering in phases $(iii-iv)$, for large \Peclet numbers, depends on the ratio $Pe/\Gamma^{1.6}$ exactly as for the topological model, see Fig. \ref{fig:collapse2} $(a)$.
Here however, in the clustering phases $(iii-iv)$ the clusters are larger not only containing $\approx M$ particles but $\approx 2000$ from $N=10000$.
For the other transition, from the clustering phases $(iii-iv)$ towards the percolating patchy state $(v)$ at small \Peclet numbers, the exact value of the order parameter $\langle \Phi_c \rangle$ seems to depend on both $Pe$ and $\Gamma$, however, the onset point of clustering (that is the point where $\langle \Phi_c \rangle$ starts to decrease) depends only on the ration $Pe/\Gamma$, see Fig. \ref{fig:collapse2} $(b)$.
Despite the neighborhood network, that only serves as a proper indicator of clustering when the density is chosen properly, we can also look at local correlations by introducing the following integral of the radial distribution function on the length scale of the interaction radius
\begin{align}
	C_2:=\rho^2\int_{\mathbb{R}^2} d \mathbf{r}_1 d \mathbf{r_2} [g(|\mathbf{r}_2-\mathbf{r}_1|)-1]\theta(R-|\mathbf{r}_1|)\theta(R-|\mathbf{r}_2|),
	\label{eq:def_C2}
\end{align}
which was employed already in previous studies \cite{\Ctwo}.
Here, $\theta$ denotes the Heaviside function and $g(r)$ the radial distribution function.
In the rescaled coordinates that we are employing, the interaction radius is always $R=1$.
The meaning of the parameter $C_2$ can be intuitively understood as follows.
Assuming an independent and homogeneous distribution of all particle positions, the probability of finding particles one and two simultaneously in an arbitrary unit circle is (in the thermodynamic limit) $(M/N)^2$.
If nonzero pair correlations are considered, then this probability is enlarged to $(M/N)^2 +C_2/N^2$.
Thus, once $C_2$ reaches $M^2$, the probability of finding two particular particles in the same unit circle is doubled compared to mean field predictions.
In that way $C_2$ serves as an order parameter of clustering on the length scale of the interaction length $R$.

In Fig. \ref{fig:collapse2} $(c)$ we display $C_2$ as a function of $Pe/\Gamma^{1.6}$ for various, high, values of $Pe$ for the transition between phases $(i-ii)$ and $(iii-iv)$.
This order parameter confirms that clustering only depends on the ratio $Pe/\Gamma^{1.6}$ in the high \Peclet limit.
In Fig. \ref{fig:collapse2} $(d)$ we show $C_2$ as a function of $Pe/\Gamma$ for various, small, values of $Pe$ for the transition between phases $(iii-iv)$ and $(v)$, confirming that the onset of clustering depends only on the ratio $Pe/\Gamma$.

In summary, we studied $N=10^4$ particles in a square domain with periodic boundary conditions by means of agent-based simulations.
Depending on \Peclet number and coupling strength we find various phases $(i-vi)$ that are characterized by orientational order and spatial distribution.
In terms of orientation we find either global polarization of the self-propulsion direction or orientational disorder.
In terms of spatial distribution we find either homogeneous states, band states (that have been intensively studied before and are not discussed here) or clustering states.
By clustering states we mean local accumulations of small groups of particles that interact with each other.
We focused on a detailed study of clustering that seems to be independent of polarization.
There is strong clustering in phases $(iii-iv)$ and no or only little clustering in the remaining phases.
The transition from $(i-ii)$ to $(iii-iv)$ at high \Peclet numbers depends on the ratio $Pe/\Gamma^{\alpha}$ as previously suggested for topological neighborhoods \cite{ZHR22}.
We do find the same relation with $\alpha=1.6$ for both, metric and topological neighborhoods.
The transition from $(iii-iv)$ to $(v)$ at small \Peclet numbers also depends on the ratio $Pe/\Gamma^{\alpha}$ but with $\alpha=1.0$ for both, metric and topological neighborhoods.

\section{Formulation of the Repeated Ringkinetic Theory \label{sec:ringkin}}

\subsection{Topological Neighborhoods}

\subsubsection{Starting Point: $N$-particle Fokker-Planck Equation}
Equivalently to the Langevin dynamics, \cref{eq:langevin}, we can write the Fokker-Planck equation describing the time evolution of the $N$-particle probability density function
\begin{align}
	&\partial_t P_N= -Pe \sum_{i=1}^{N} ( \cos \phi_i \partial_{x_i}  + \sin \phi_i \partial_{y_i} ) P_N
	-\Gamma \sum_{i=1}^{N}\partial_{\phi_i} \sum_{j\in \Omega_i} K(\phi_j-\phi_i)P_N + \sum_{i=1}^{N}\frac{1}{2}\partial_{\phi_i}^2P_N,
	\label{eq:NFP}
\end{align}
where the $N$-particle probability $P_N$ depends on all positions and orientations $\{x_i, y_i, \phi_i\}_{i\in 1, \dots, N}$.
We can rewrite the sum over $j$ in \cref{eq:NFP} by introducing the indicator function
\begin{align}
	\mathbbm{1}_{j\in\Omega_i}=\begin{cases} 1 \text{ if } j \in \Omega_i \\ 0 \text{ else} \end{cases}
	\label{eq:indicator_function}
\end{align}
to arrive at
\begin{align}
	&\partial_t P_N= -Pe \sum_{i=1}^{N} ( \cos \phi_i \partial_{x_i}  + \sin \phi_i \partial_{y_i} ) P_N
	-\Gamma \sum_{i,j=1}^{N}\partial_{\phi_i} \mathbbm{1}_{j\in\Omega_i} K(\phi_j-\phi_i)P_N + \sum_{i=1}^{N}\frac{1}{2}\partial_{\phi_i}^2P_N.
	\label{eq:NFP_indicator}
\end{align}
Note that the indicator function $\mathbbm{1}_{j\in\Omega_i}$ depends on the position of all particles $\{\mathbf{r}_i\}_{i\in 1, \dots, N}$.

\subsubsection{Symmetry}
We assume that we can not statistically distinguish different particles, that is we assume that $P_N$ is symmetric with respect to permutations of the particles
\begin{align}
	&P_N(\mathbf{r}_1, \phi_1, \mathbf{r}_2, \phi_2, \dots, \mathbf{r}_N, \phi_N)
	=P_N(\mathbf{r}_{\pi(1)}, \phi_{\pi(1)}, \mathbf{r}_{\pi(2)}, \phi_{\pi(2)}, \dots, \mathbf{r}_{\pi(N)}, \phi_{\pi(N)})
	\label{eq:permutation_symmetry}
\end{align}
for any permutation $\pi$ of the indexes $1, 2, \dots, N$.

\subsubsection{Integrating Degrees of Freedom not explicitly considered}

In general, we can reduce the $N$-particle probability towards an one- or two-particle probability by integrating over all remaining degrees of freedom
\begin{align}
	&P_1(\mathbf{r}_1, \phi_1)= \int d \mathbf{r}_2 \dots d \mathbf{r}_N d \phi_2 \dots d \phi_N P_N,
	\label{eq:reduction1}
	\\
	&P_2(\mathbf{r}_1, \phi_1, \mathbf{r}_2, \phi_2)= \int d \mathbf{r}_3 \dots d \mathbf{r}_N d \phi_3 \dots d \phi_N P_N.
	\label{eq:reduction2}
\end{align}
Deriving \cref{eq:reduction1,eq:reduction2} with respect to time and inserting \cref{eq:NFP_indicator} we obtain
\begin{align}
	&\partial_t P_1(\mathbf{r}_1, \phi_1)=-Pe ( \cos \phi_1 \partial_{x_1}  + \sin \phi_1 \partial_{y_1} ) P_1+ \frac{1}{2}\partial_{\phi_1}^2P_1
	-\int d \mathbf{r}_2 \dots d \mathbf{r}_N d \phi_2 \dots d \phi_N \Gamma \sum_{j=2}^{N}\partial_{\phi_1} \mathbbm{1}_{j\in\Omega_1} K(\phi_j-\phi_1)P_N ,
	\label{eq:1FP}
	\\
	&\partial_t P_2(\mathbf{r}_1, \phi_1, \mathbf{r}_2, \phi_2)=-Pe ( \cos \phi_1 \partial_{x_1}  + \sin \phi_1 \partial_{y_1} ) P_2
	-Pe ( \cos \phi_2 \partial_{x_2}  + \sin \phi_2 \partial_{y_2} ) P_2+ \frac{1}{2}(\partial_{\phi_1}^2+\partial_{\phi_2}^2)P_2
	\notag
	\\
	&-\int d \mathbf{r}_3 \dots d \mathbf{r}_N d \phi_3 \dots d \phi_N \Gamma \partial_{\phi_1} \mathbbm{1}_{2\in\Omega_1} K(\phi_2-\phi_1)P_N 
	-\int d \mathbf{r}_3 \dots d \mathbf{r}_N d \phi_3 \dots d \phi_N \Gamma \partial_{\phi_2} \mathbbm{1}_{1\in\Omega_2} K(\phi_1-\phi_2)P_N 
	\notag
	\\
	&-\int d \mathbf{r}_3 \dots d \mathbf{r}_N d \phi_3 \dots d \phi_N \Gamma \sum_{j=3}^{N}\partial_{\phi_1} \mathbbm{1}_{j\in\Omega_1} K(\phi_j-\phi_1)P_N
	-\int d \mathbf{r}_3 \dots d \mathbf{r}_N d \phi_3 \dots d \phi_N \Gamma \sum_{j=3}^{N}\partial_{\phi_2} \mathbbm{1}_{j\in\Omega_2} K(\phi_j-\phi_2)P_N ,
	\label{eq:2FP}
\end{align}
where some terms canceled due to the periodic boundary conditions.
Using the symmetry \cref{eq:permutation_symmetry} we can replace all summands in the sums over $j$ by just one representative term to arrive at
\begin{align}
	&\partial_t P_1(\mathbf{r}_1, \phi_1)=-Pe ( \cos \phi_1 \partial_{x_1}  + \sin \phi_1 \partial_{y_1} ) P_1+ \frac{1}{2}\partial_{\phi_1}^2P_1
	-(N-1)\int d \mathbf{r}_2 \dots d \mathbf{r}_N d \phi_2 \dots d \phi_N \Gamma \partial_{\phi_1} \mathbbm{1}_{2\in\Omega_1} K(\phi_2-\phi_1)P_N ,
	\label{eq:1FP2}
	\\
	&\partial_t P_2(\mathbf{r}_1, \phi_1, \mathbf{r}_2, \phi_2)=-Pe ( \cos \phi_1 \partial_{x_1}  + \sin \phi_1 \partial_{y_1} ) P_2
	-Pe ( \cos \phi_2 \partial_{x_2}  + \sin \phi_2 \partial_{y_2} ) P_2+ \frac{1}{2}(\partial_{\phi_1}^2+\partial_{\phi_2}^2)P_2
	\notag
	\\
	&-\int d \mathbf{r}_3 \dots d \mathbf{r}_N d \phi_3 \dots d \phi_N \Gamma \partial_{\phi_1} \mathbbm{1}_{2\in\Omega_1} K(\phi_2-\phi_1)P_N 
	-\int d \mathbf{r}_3 \dots d \mathbf{r}_N d \phi_3 \dots d \phi_N \Gamma \partial_{\phi_2} \mathbbm{1}_{1\in\Omega_2} K(\phi_1-\phi_2)P_N 
	\notag
	\\
	&-(N-2)\int d \mathbf{r}_3 \dots d \mathbf{r}_N d \phi_3 \dots d \phi_N \Gamma \partial_{\phi_1} \mathbbm{1}_{3\in\Omega_1} K(\phi_3-\phi_1)P_N
	-(N-2)\int d \mathbf{r}_3 \dots d \mathbf{r}_N d \phi_3 \dots d \phi_N \Gamma \partial_{\phi_2} \mathbbm{1}_{3\in\Omega_2} K(\phi_3-\phi_2)P_N .
	\label{eq:2FP2}
\end{align}
So far \cref{eq:1FP2,eq:2FP2} are exact.
One can write down analogous equations for $P_3$, $P_4$, etc..
However, in this paper we are considering only $P_1$ and $P_2$.
Note that the right hand side of \cref{eq:1FP2,eq:2FP2} depends on the full $N$-particle probability (at least the spatial part).
This is due to the complicated indicator functions $\mathbbm{1}_{2\in \Omega_1}$, $\mathbbm{1}_{3\in \Omega_1}$ and $\mathbbm{1}_{3\in \Omega_2}$ that depend on the positions of all particles.
In contrast, for additive pairwise interactions, there would be only $P_2$ and $P_3$ on the right hand side of \cref{eq:1FP2,eq:2FP2}, respectively.

\subsubsection{Ursell Expansion}
The $N$-particle probability $P_N$ can be expanded systematically in a sum of products of functions depending only on the degrees of freedom of one, two, three, etc. particles by means of the Ursell expansion \cite{\ursell}.
For simplicity we introduce the following short notation. 
We denote all degrees of freedom of particles one, two, three, etc. by $1$, $2$, $3$, $\dots$, respectively.
That means in the present case $1=(x_1, y_1, \phi_1)$ and by the integral $\int d 1$ we mean $\int_0^{L_x}d x_1\int_0^{L_y}d y_1 \int_{-\pi}^{\pi}d \phi_1$.
We denote the set of decompositions of $\{1, 2, \dots, k\}$ into unordered tuplets exept the trivial $k$-plet $(1, 2, \dots, k)$ itself by $\mathcal{T}_k$.
That means, for example
\begin{align}
	&\mathcal{T}_2=\{\{1, 2\}\},
	\label{eq:duplets}
	\\
	&\mathcal{T}_3=\{\{1, 2, 3\}, \{(1, 2),3\}, \{(1, 3),2\}, \{(2, 3),1\}\}.
	\label{eq:triplets}
\end{align}
We define the correlation functions $G_1(1)$, $G_2(1,2)$, $G_3(1,2,3)$, etc. recursively via
\begin{align}
	&G_1(1):=P_1(1),
	\label{eq:Gone}
	\\
	&G_k(1, 2, \dots, k):=P_k(1, 2, \dots, k)-\sum_{\tau\in \mathcal{T}_k}\prod_{s\in \tau}G_{|s|}(s),
	\label{eq:Gk}
\end{align}
where $|s|$ denotes the dimension of the tuplet.
Following this definition, the first few $G$-functions are
\begin{align}
	&G_1(1):= P_1(1),
	\label{eq:G1}
	\\
	&G_2(1,2):=P_2(1,2)-P_1(1)P_1(2),
	\label{eq:G2}
	\\
	&G_3(1,2,3):=P_3(1,2,3)-G_2(1,2)P_1(3)
	-G_2(1,3)P_1(2)
	-G_2(2,3)P_1(1)-P_1(1)P_1(2)P_1(3).
	\label{eq:G3}
\end{align}
Note that due to the symmetry of $P_k$ with respect to permutations of particles, \cref{eq:permutation_symmetry}, we have the same symmetry for $G_k$, that means $G_2(1,2)=G_2(2,1)$, etc..
The function $G_1$ describes the one particle probability.
Thus it is not really a correlation function and we will prefarably use the notion $P_1$ instead of $G_1$.
However, for $k>1$, $G_k$ really describes correlations.
In particular for $G_2 \equiv G_3, \equiv G_4 \equiv \dots \equiv 0$ all particles are independent.
This is the typical assumption of mean field theories.
One important property of the $G$-functions for $k>1$ is
\begin{align}
	\int d 1 G_k(1, 2, \dots)=0.
	\label{eq:integral_gfunction}
\end{align}
That means the integral of a correlation function over all degrees of freedom of one particle (no matter which, due to the permutation symmetry) vanishes.
For $k=2$ this property can be seen immediatly by integrating \cref{eq:G2} over $1$.
It is straightforward to show \cref{eq:integral_gfunction} for $k>2$ inductively.

\subsubsection{Repeated Ringkinetic Approach}

In mean field theories all correlation functions $G_2$, $G_3$, etc. are assumed to be zero.
In the ringkinetic approach one assumes that $G_k\equiv 0$ for all $k\ge 3$ and one sets $G_2\equiv 0$ on the right hand side of \cref{eq:2FP2} in order to calculate $G_2$ and use it on the right hand side of \cref{eq:1FP2} for the dynamics of $P_1$.

In contrast, in the repeated ringkinetic approach, nonzero $G_2$ is considered in all equations.
In particular the impact of $G_2$ on $G_2$ itself is considered.
We are following this aproach in the present paper.
That means the only approximation used is that correlations between three or more particles are negligible.
It should be noted that the approximation
\begin{align}
	G_3\equiv G_4\equiv \dots \equiv 0
	\label{eq:ringkin_approximation}
\end{align}
is neither violating the symmetry \cref{eq:permutation_symmetry} nor any other fundamental physical principle.
In fact, there are mathematically valid $N$-particle probabilities that satisfy $G_2\not\equiv 0$ and \cref{eq:ringkin_approximation} exactly.
One can easily construct an example in the following way: Start with the probability of $N$ independent, identically distributed particles and perform one iteration of an arbitrary Markov-chain that respects the symmetry \cref{eq:permutation_symmetry} and that depends only on interactions between pairs of particles (no three particle interactions). 
In this way, in the first step, only $G_2$ is generated. 
After one more step $G_3$ is generated from $G_2$, etc..
However, the point raised here, is only that the assumtion \cref{eq:ringkin_approximation} is not violating any fundamental principles of probability.
For experimental interacting many particle systems \cref{eq:ringkin_approximation} is usually not valid exactly, however, it might still be a reasonable approximation and certainly it is an improvement over mean field theory.
Throughout the rest of the paper we assume \cref{eq:ringkin_approximation} without explicitly mentioning it each time.

We consider a function $h(1, 2, 3, \dots, N)$ that can depend on the state of all particles. 
We assume furthermore that $h$ is invariant under arbitrary permutations of the particles $\{3, 4, 5, \dots, N\}$.
Then
\begin{align}
	 &h(1, \dots) P_N(1, \dots) 
	= h(1, \dots) P_1(1) P_1(2) P_{N-2}(3, \dots) 
	+ h(1, \dots) G_2(1,2) P_{N-2}(3, \dots)
	\notag
	\\
	&+ h(1, \dots)  \sum_{i=3}^N G_2(1,i) P_1(2) P_{N-3}(\{3, \dots\}/\{i\})
	+ h(1, \dots) \sum_{i=3}^N G_2(2,i) P_1(1) P_{N-3}(\{3, \dots\}/\{i\})
	\notag
	\\
	&+ h(1, \dots) \sum_{i, j=3, i\neq j}^N G_2(1,i) G_2(2, j) P_{N-4}(\{3, \dots\}/\{i, j\})
	\label{eq:Ursell_P3_1}
	\\
	&= h(1, \dots) P_1(1) P_1(2) P_{N-2}(3, \dots)
	+ h(1, \dots) G_2(1,2) P_{N-2}(3, \dots)
	+(N-2) h(1, \dots)  G_2(1,3) P_1(2) P_{N-3}(\{4, \dots\}) 
	\notag
	\\
	&+(N-2) h(1, \dots) G_2(2,3) P_1(1) P_{N-3}(\{4, \dots\})
	+(N-2)(N-3) h(1, \dots) G_2(1,3) G_2(2, 4) P_{N-4}(\{5, \dots\}).
	\label{eq:Ursell_P3_2}
\end{align}
For a proof of \cref{eq:Ursell_P3_1} we just insert the Ursell expansion \cref{eq:Gone,eq:Gk} for $P_N$, $P_{N-2}$, $P_{N-3}$ and $P_{N-4}$.
Exploiting the permutation symmetry \cref{eq:permutation_symmetry} we can replace the sums over $i$ and $j$ by just one representative term for each sum, arriving at \cref{eq:Ursell_P3_2}.
Analogously, if $h(1, \cdots, N)$ is invariant under permutations of $\{4, \cdots, N\}$, we find
\begin{align}
	&h(1, \dots)P_N(1, \dots)=h(1, \dots)P_1(1)P_1(2)P_1(3)P_{N-3}(4, \dots)+h(1, \dots)G_2(1,2)P_1(3)P_{N-3}(4, \dots)
	\notag
	\\
	&+h(1, \dots)G_2(1,3)P_1(2)P_{N-3}(4, \dots)+h(1, \dots)G_2(2,3)P_1(1)P_{N-3}(4, \dots)+(N-3)h(1, \dots)G_2(1,4)P_1(2)P_1(3)P_{N-4}(5, \dots)
	\notag
	\\
	&+(N-3)h(1, \dots)G_2(1,4)G_2(2,3)P_{N-4}(5, \dots)+(N-3)h(1, \dots)G_2(2,4)P_1(1)P_1(3)P_{N-4}(5, \dots)
	\notag
	\\
	&+(N-3)h(1, \dots)G_2(2,4)G_2(1,3)P_{N-4}(5, \dots)
	+(N-3)h(1, \dots)G_2(3,4)P_1(2)P_1(2)P_{N-4}(5, \dots)
	\notag
	\\
	&+(N-3)h(1, \dots)G_2(3,4)G_2(1,2)P_{N-4}(5, \dots)+(N-3)(N-4)h(1, \dots)G_2(1,4)G_2(2, 5)P_1(3)P_{N-5}(6, \dots)
	\notag
	\\
	&+(N-3)(N-4)h(1, \dots)G_2(1,4)G_2(3, 5)P_1(2)P_{N-5}(6, \dots)+(N-3)(N-4)h(1, \dots)G_2(2,4)G_2(3, 5)P_1(1)P_{N-5}(6, \dots)
	\notag
	\\
	&+(N-3)(N-4)(N-5)h(1, \dots)G_2(1,4)G_2(2, 5)G_2(3,6)P_{N-6}(7, \dots).
	\label{eq:Ursell_P4}
\end{align}

Note that in \cref{eq:1FP2,eq:2FP2} there appear terms of the types of \cref{eq:Ursell_P3_2,eq:Ursell_P4} with $h$-function
\begin{align}
	h(1, 2, \dots, N)= \begin{cases} \mathbbm{1}_{2\in \Omega_1}(\mathbf{r}_1, \mathbf{r}_2, \mathbf{r_3}, \dots, \mathbf{r}_N) \text{ in \cref{eq:1FP2} and first two interaction terms in \cref{eq:2FP2}}\\
\mathbbm{1}_{3\in \Omega_1}(\mathbf{r}_1, \mathbf{r}_2, \mathbf{r_3}, \dots, \mathbf{r}_N) \text{ in third interaction term in \cref{eq:2FP2}}
\\
\mathbbm{1}_{3\in \Omega_2}(\mathbf{r}_1, \mathbf{r}_2, \mathbf{r_3}, \dots, \mathbf{r}_N) \text{ in fourth interaction term in \cref{eq:2FP2}.}
	\end{cases}
	\label{eq:h-function}
\end{align}
In the following we evaluate a few integrals that are required to simplify \cref{eq:1FP2,eq:2FP2}.
We denote the expectation value of the $h$-function $h=\mathbbm{1}_{2\in\Omega_1}$ with respect to $P_{N-2}(3,\dots, N)$ by
\begin{align}
	q^{N}_{k}(\mathbf{r}_1, r_{12}):=\int \mathbbm{1}_{2\in \Omega_1}(\mathbf{r}_1, \mathbf{r}_2, \mathbf{r_3}, \dots, \mathbf{r}_N) P_{N-2}(3, 4, \ldots, N) d 3 \dots d N,
	\label{eq:neighbor_probability_N}
\end{align}
where $r_{ij}:=|\mathbf{r}_i-\mathbf{r}_j|$ denotes the distance between particles $i$ and $j$.
We consider the thermodynamic limit for $\frac{N}{L_xL_y}=const.$
\begin{align}
	q_{k}(\mathbf{r}_1, r_{12})= \lim_{N\rightarrow \infty} q^{N}_{k}(\mathbf{r}_1, r_{12}).
	\label{eq:neighbor_probability}
\end{align}
It describes the probability that there are less than $k$ particles (other than particles one and two) that are closer to particle $1$ than the distance between particles one and two under the assumption that the particles $\{3, \cdots, N\}$ are independent on particles one and two in the limit of large system size.
Obviously, this probability depends on the position of particle one and the distance between particles one and two, furthermore it depends on the distribution $P_{N-2}$.
We denote the probability that there are exactly $s$ particles that are closer to particle one than the distance between particles one and two by $p_s$, thus
\begin{align}
	q_{k}(\mathbf{r}_1, r_{12})= \sum_{s=0}^{k-1} p_{s}(\mathbf{r}_1, r_{12}).
	\label{eq:neighbor_probability2}
\end{align}
Recently, $p_s$ was calculated exactly for particles correlated up to arbitrary, but finite order \cite{\correlations}.
Under the assumption of \cref{eq:ringkin_approximation} the result depends only on two integrals:
\begin{align}
	&C_1(\mathbf{r}_1, r_{12}):= N \int d 3 P_1(3) \theta(r_{12}-r_{13}),
	\label{eq:C1}
	\\
	&C_2(\mathbf{r}_1, r_{12}):= N^2 \int d 3 d 4 G_2(3,4) \theta(r_{12}-r_{13})\theta(r_{12}-r_{14}),
	\label{eq:C2}
\end{align}
where $\theta$ denotes the Heaviside-function.
With this definition we have \cite{\correlations}
\begin{align}
	p_s(\mathbf{r}_1, r_{12})= &(C_1-C_2)^{s} \exp(C_2/2-C_1) \sum_{q=0}^{\infty} \bigg[ \frac{C_2}{2(C_1-C_2)^2}\bigg]^{q} \frac{1}{q! (s-2q)!}
	\label{eq:KSZIresult}
\end{align}
with the convention that $1/l!:=0$ for $l<0$ such that there are only finitely many nonzero terms in \cref{eq:KSZIresult}.
Note that this result was derived in Refs. \cite{\correlations} for spatially homogeneous systems, such that $C_{1/2}$ is not explicitly depending on the position $\mathbf{r}_1$.
However, this assumption is not needed. 
One can explicitly go through all steps in the derivation of \cref{eq:KSZIresult} in Refs. \cite{\correlations} without any change when the assumption of spatial homogeneity is dropped.

Similar to \cref{eq:neighbor_probability_N,eq:neighbor_probability} we find
\begin{align}
	[1-\theta(r_{13}-r_{12})]q_{k-1}(\mathbf{r}_1, r_{12})&=\lim_{N\rightarrow \infty}\int \mathbbm{1}_{2\in \Omega_1}(\mathbf{r}_1, \ldots, \mathbf{r}_N)[1-\theta(r_{13}-r_{12})]P_{N-3}(4, \ldots, N) d4 \ldots dN,
	\label{eq:qk_PN3_1}
	\\
	\theta(r_{13}-r_{12})q_{k}(\mathbf{r}_1, r_{12})&=\lim_{N\rightarrow \infty}\int \mathbbm{1}_{2\in \Omega_1}(\mathbf{r}_1, \ldots, \mathbf{r}_N)\theta(r_{13}-r_{12})P_{N-3}(4, \ldots, N) d4 \ldots dN,
	\label{eq:qk_PN3_2}
\end{align}
where the expectation value is taken only with respect to particles four to $N$ which does not change anything essential in the thermodynamic limit.
However, the resulting expectation value does depend on the position of particle three: \cref{eq:qk_PN3_1} considers the case that the distance between particles one and three is larger than the distance between particles one and two, whereas \cref{eq:qk_PN3_2} considers the opposite case that particle three is further away from particle one than particle two.
Summing \cref{eq:qk_PN3_1,eq:qk_PN3_2} results in
\begin{align}
	\lim_{N\rightarrow \infty}\int \mathbbm{1}_{2\in\Omega_1}(\mathbf{r}_1, \ldots, \mathbf{r}_N)P_{N-3}(4, \ldots, N)d4 \ldots dN
	=\theta(r_{13}-r_{12})q_k(\mathbf{r}_1, r_{12}) + [1-\theta(r_{13}-r_{12})]q_{k-1}(\mathbf{r}_1, r_{12}).
	\label{eq:qk_PN3}
\end{align}
Analogously, we find
\begin{align}
	&\lim_{N\rightarrow \infty}\int \mathbbm{1}_{2\in\Omega_1}(\mathbf{r}_1, \ldots, \mathbf{r}_N)P_{N-4}(5, \ldots, N)d4 \ldots dN
	=\theta(r_{13}-r_{12}) \theta(r_{14}-r_{12})q_k(\mathbf{r}_1, r_{12})
	\notag
	\\&+ \big\{\theta(r_{13}-r_{12})[1-\theta(r_{14}-r_{12})]+[1-\theta(r_{13}-r_{12})]\theta(r_{14}-r_{12})\big\}q_{k-1}(\mathbf{r}_1, r_{12})
	\notag
	\\&+[1-\theta(r_{13}-r_{12})][1- \theta(r_{14}-r_{12})]q_{k-2}(\mathbf{r}_1, r_{12}),
	\label{eq:qk_PN4}
\end{align}
\begin{align}
	&\lim_{N\rightarrow \infty}\int \mathbbm{1}_{2\in\Omega_1}(\mathbf{r}_1, \ldots, \mathbf{r}_N)P_{N-5}(6, \ldots, N)d4 \ldots dN
	=\theta(r_{13}-r_{12}) \theta(r_{14}-r_{12})\theta(r_{15}-r_{12})q_k(\mathbf{r}_1, r_{12})
	\notag
	\\&+ \big\{[1-\theta(r_{13}-r_{12})]\theta(r_{14}-r_{12})\theta(r_{15}-r_{12})+\theta(r_{13}-r_{12})[1-\theta(r_{14}-r_{12})]\theta(r_{15}-r_{12})
	\notag
	\\
	&+\theta(r_{13}-r_{12})\theta(r_{14}-r_{12})[1-\theta(r_{15}-r_{12})]\big\}q_{k-1}(\mathbf{r}_1, r_{12})
	+\big\{[1-\theta(r_{13}-r_{12})][1- \theta(r_{14}-r_{12})]\theta(r_{15}-r_{12})
	\notag
	\\
	&+[1-\theta(r_{13}-r_{12})]\theta(r_{14}-r_{12})[1- \theta(r_{15}-r_{12})]+\theta(r_{13}-r_{12})[1- \theta(r_{14}-r_{12})][1- \theta(r_{15}-r_{12})]\big\}q_{k-2}(\mathbf{r}_1, r_{12})
	\notag
	\\
	&+[1-\theta(r_{13}-r_{12})][1-\theta(r_{14}-r_{12})][1- \theta(r_{15}-r_{12})]q_{k-3}(\mathbf{r}_1, r_{12}),
	\label{eq:qk_PN5}
\end{align}
and
\begin{align}
	&\lim_{N\rightarrow \infty}\int \mathbbm{1}_{2\in\Omega_1}(\mathbf{r}_1, \ldots, \mathbf{r}_N)P_{N-6}(7, \ldots, N)d4 \ldots dN
	=\theta(r_{13}-r_{12}) \theta(r_{14}-r_{12})\theta(r_{15}-r_{12})\theta(r_{16}-r_{12})q_k(\mathbf{r}_1, r_{12})
	\notag
	\\
	&+ \big\{[1-\theta(r_{13}-r_{12})]\theta(r_{14}-r_{12})\theta(r_{15}-r_{12})\theta(r_{16}-r_{12})+\theta(r_{13}-r_{12})[1-\theta(r_{14}-r_{12})]\theta(r_{15}-r_{12})\theta(r_{16}-r_{12})
	\notag
	\\
	&+\theta(r_{13}-r_{12})\theta(r_{14}-r_{12})[1-\theta(r_{15}-r_{12})]\theta(r_{16}-r_{12})+\theta(r_{13}-r_{12})\theta(r_{14}-r_{12})\theta(r_{15}-r_{12})[1-\theta(r_{16}-r_{12})]\big\}q_{k-1}(\mathbf{r}_1, r_{12})
	\notag
	\\
	&+\big\{[1-\theta(r_{13}-r_{12})][1- \theta(r_{14}-r_{12})]\theta(r_{15}-r_{12})\theta(r_{16}-r_{12})+[1-\theta(r_{13}-r_{12})]\theta(r_{14}-r_{12})[1-\theta(r_{15}-r_{12})]\theta(r_{16}-r_{12})
	\notag
	\\
	&+[1-\theta(r_{13}-r_{12})]\theta(r_{14}-r_{12})\theta(r_{15}-r_{12})[1-\theta(r_{16}-r_{12})]+\theta(r_{13}-r_{12})[1- \theta(r_{14}-r_{12})][1-\theta(r_{15}-r_{12})]\theta(r_{16}-r_{12})
	\notag
	\\
	&+\theta(r_{13}-r_{12})[1- \theta(r_{14}-r_{12})]\theta(r_{15}-r_{12})[1-\theta(r_{16}-r_{12})]+\theta(r_{13}-r_{12})\theta(r_{14}-r_{12})[1-\theta(r_{15}-r_{12})][1-\theta(r_{16}-r_{12})]\big\}
	\notag
	\\
	&\times q_{k-2}(\mathbf{r}_1, r_{12})+\big\{[1-\theta(r_{13}-r_{12})][1- \theta(r_{14}-r_{12})][1-\theta(r_{15}-r_{12})]\theta(r_{16}-r_{12})
	\notag
	\\
	&+[1-\theta(r_{13}-r_{12})][1- \theta(r_{14}-r_{12})]\theta(r_{15}-r_{12})[1-\theta(r_{16}-r_{12})]+[1-\theta(r_{13}-r_{12})]\theta(r_{14}-r_{12})[1-\theta(r_{15}-r_{12})]
	\notag
	\\
	&\times[1-\theta(r_{16}-r_{12})] +\theta(r_{13}-r_{12})[1- \theta(r_{14}-r_{12})][1-\theta(r_{15}-r_{12})][1-\theta(r_{16}-r_{12})]\big\}q_{k-3}(\mathbf{r}_1, r_{12})
	\notag
	\\
	&+[1-\theta(r_{13}-r_{12})][1-\theta(r_{14}-r_{12})][1- \theta(r_{15}-r_{12})][1- \theta(r_{16}-r_{12})]q_{k-4}(\mathbf{r}_1, r_{12}).
	\label{eq:qk_PN6}
\end{align}
Considering the leading order terms for $N\rightarrow \infty$, we plug \cref{eq:Ursell_P3_2} with $h$-function according to \eqref{eq:h-function} into the interaction term in the $P_1$-equation \eqref{eq:1FP2}, we employ \cref{eq:neighbor_probability_N,eq:neighbor_probability,eq:qk_PN3,eq:qk_PN4} and we identify identical appearing integrals using \cref{eq:integral_gfunction} to arrive at
\begin{align}
	&\partial_t P_1(\mathbf{r}_1, \phi_1)=-Pe ( \cos \phi_1 \partial_{x_1}  + \sin \phi_1 \partial_{y_1} ) P_1(1)+ \frac{1}{2}\partial_{\phi_1}^2P_1(1)
	\notag
	\\
	&-N \Gamma \int d 2 \partial_{\phi_1} K(\phi_2-\phi_1) 
	P_1(1)P_1(2) q_k(\mathbf{r}_1, r_{12})
	-N \Gamma \int d 2 \partial_{\phi_1} K(\phi_2-\phi_1) 
	G_2(1,2) q_k(\mathbf{r}_1, r_{12})
	\notag
	\\
	&-N^{2} \Gamma \int d 2 d 3 \partial_{\phi_1} K(\phi_2-\phi_1)\theta(r_{12}-r_{13})
	G_2(1,3) P_1(2) [q_{k-1}(\mathbf{r}_1, r_{12}) - q_{k}(\mathbf{r}_1, r_{12})]
	\notag
	\\
	&- N^{2} \Gamma \int d 2 d 3 \partial_{\phi_1} K(\phi_2-\phi_1)\theta(r_{12}-r_{13})
	G_2(2,3) P_1(1) [q_{k-1}(\mathbf{r}_1, r_{12}) - q_{k}(\mathbf{r}_1, r_{12})]
	\notag
	\\
	&- N^{3} \Gamma \int d 2 d 3 d 4 \partial_{\phi_1} K(\phi_2-\phi_1)\theta(r_{12}-r_{13})\theta(r_{12}-r_{14})
	 G_2(1,3) G_2(2,4)
	 [q_{k-2}(\mathbf{r}_1, r_{12}) 
	 - 2 q_{k-1}(\mathbf{r}_1, r_{12})+  q_{k}(\mathbf{r}_1, r_{12})].
		\label{eq:BBGKY1}
\end{align}
Thus, we found a time evolution equation for the one-particle-distribution $P_1$ that only depends on the one-particle-distribution itself and on the pair correlation function $G_2$.
However, the equation is highly nonlinear as can be seen from the appearance of terms like $P_1 P_1$ or $G_2 G_2$, etc. in \cref{eq:BBGKY1}.
Furthermore, the $q_k$-terms appearing in \cref{eq:BBGKY1} are function-valued nonlinear functionals of $P_1$ and $G_2$, see \cref{eq:neighbor_probability2,eq:C1,eq:C2,eq:KSZIresult}.

To close the hierarchy, we need the time evolution equation of the pair correlation function $G_2$.
We start writing the time evolution equation for $P_2$ that can be obtained in full analogy to \cref{eq:BBGKY1}.
Considering the leading order terms for large $N$ we start from \cref{eq:2FP2}, plug in \cref{eq:Ursell_P3_2,eq:Ursell_P4}for the interaction integrals in \cref{eq:2FP2} with a proper choice of the $h$-function according to \cref{eq:h-function}.
Next, we perform the integrals over degrees of freedoms that are not involved in the interaction according to \cref{eq:neighbor_probability_N,eq:neighbor_probability,eq:qk_PN3,eq:qk_PN4,eq:qk_PN5,eq:qk_PN6}.
Eventually we identify identical integrals employing \cref{eq:integral_gfunction} to arrive at
\begin{align}
	&\partial_t P_2(1,2)=[-Pe (\cos \phi_1\partial_{x_1}+\sin \phi_1\partial_{y_1}
	)+\frac{1}{2}\partial_{\phi_1}^2
	][P_1(1)P_1(2)+G_2(1,2)]
	\notag
	\\
	&-\Gamma q_k(\mathbf{r}_1, r_{12})\partial_{\phi_1}K(\phi_2-\phi_1)[P_1(1)P_1(2)+G_2(1,2)]
	\notag
	\\
	&-N\Gamma[q_{k-1}(\mathbf{r}_1, r_{12})-q_{k}(\mathbf{r}_1, r_{12})]P_1(2)\partial_{\phi_1}K(\phi_2-\phi_1)\int d 3 G_2(1,3)\theta(r_{12}-r_{13})
	\notag
	\\
	&-N\Gamma[q_{k-1}(\mathbf{r}_1, r_{12})-q_{k}(\mathbf{r}_1, r_{12})]\partial_{\phi_1}K(\phi_2-\phi_1)P_1(1)\int d 3 G_2(2,3) \theta(r_{12}-r_{13})
	\notag
	\\
	&-N^2\Gamma[q_{k-2}(\mathbf{r}_1, r_{12})-2q_{k-1}(\mathbf{r}_1, r_{12})+q_{k}(\mathbf{r}_1, r_{12})]\partial_{\phi_1}K(\phi_2-\phi_1)\int d 3 d 4 G_2(1,3)G_2(2,4)\theta(r_{12}-r_{13})\theta(r_{12}-r_{14})
	\notag
	\\
	&-N\Gamma \partial_{\phi_1}\int d 3 K(\phi_3-\phi_1)q_{k}(\mathbf{r}_1, r_{13})\theta(r_{12}-r_{13})[P_1(1)P_1(2)P_1(3)+G_2(1,2)P_1(3)+G_2(1,3)P_1(2)+P_1(1)G_2(2,3)]
	\notag
	\\
	&-N\Gamma \partial_{\phi_1}\int d 3 K(\phi_3-\phi_1)q_{k-1}(\mathbf{r}_1, r_{13})\theta(r_{13}-r_{12})[P_1(1)P_1(2)P_1(3)+G_2(1,2)P_1(3)+G_2(1,3)P_1(2)+P_1(1)G_2(2,3)]
	\notag
	\\
	&-N^2\Gamma \partial_{\phi_1}\int d 3 d 4 K(\phi_3-\phi_1)[q_{k-1}(\mathbf{r}_1, r_{13})-q_{k}(\mathbf{r}_1, r_{13})]\theta(r_{12}-r_{13})\theta(r_{13}-r_{14})
	\notag
	\\
	&\times[G_2(1,4)P_1(2)P_1(3)+G_2(1,4)G_2(2,3)+G_2(2,4)P_1(1)P_1(3)+G_2(2,4)G_2(1,3)+G_2(3,4)P_1(1)P_1(2)+G_2(3,4)G_2(1,2)]
	\notag
	\\
	&-N^2\Gamma \partial_{\phi_1}\int d 3 d 4 K(\phi_3-\phi_1)[q_{k-2}(\mathbf{r}_1, r_{13})-q_{k-1}(\mathbf{r}_1, r_{13})]\theta(r_{13}-r_{12})\theta(r_{13}-r_{14})
	\notag
	\\
	&\times[G_2(1,4)P_1(2)P_1(3)+G_2(1,4)G_2(2,3)+G_2(2,4)P_1(1)P_1(3)+G_2(2,4)G_2(1,3)+G_2(3,4)P_1(1)P_1(2)+G_2(3,4)G_2(1,2)]
	\notag
	\\
	&-N^3\Gamma \partial_{\phi_1}\int d 3 d 4 d 5 K(\phi_3-\phi_1)[q_{k-2}(\mathbf{r}_1, r_{13})-2q_{k-1}(\mathbf{r}_1, r_{13})+q_{k}(\mathbf{r}_1, r_{13})]\theta(r_{12}-r_{13})\theta(r_{13}-r_{14})\theta(r_{13}-r_{15})
	\notag
	\\
	&\times[G_2(1,4)G_2(2,5)P_1(3)+G_2(1,4)P_1(2)G_2(3,5)+P_1(1)G_2(2,4)G_2(3,5)]
	\notag
	\\
	&-N^3\Gamma \partial_{\phi_1}\int d 3 d 4 d 5 K(\phi_3-\phi_1)[q_{k-3}(\mathbf{r}_1, r_{13})-2q_{k-2}(\mathbf{r}_1, r_{13})+q_{k-1}(\mathbf{r}_1, r_{13})]\theta(r_{13}-r_{12})\theta(r_{13}-r_{14})\theta(r_{13}-r_{15})
	\notag
	\\
	&\times[G_2(1,4)G_2(2,5)P_1(3)+G_2(1,4)P_1(2)G_2(3,5)+P_1(1)G_2(2,4)G_2(3,5)]
	\notag
	\\
	&-N^4\Gamma \partial_{\phi_1}\int d 3 d 4 d 5 d 6 K(\phi_3-\phi_1)[q_{k-3}(\mathbf{r}_1, r_{13})-3q_{k-2}(\mathbf{r}_1, r_{13})+3q_{k-1}(\mathbf{r}_1, r_{13})-q_{k}(\mathbf{r}_1, r_{13})]\theta(r_{12}-r_{13})\theta(r_{13}-r_{14})
	\notag
	\\
	&\times\theta(r_{13}-r_{15})\theta(r_{13}-r_{16})G_2(1,4)G_2(2,5)G_2(3,6)
	\notag
	\\
	&-N^4\Gamma \partial_{\phi_1}\int d 3 d 4 d 5 d 6 K(\phi_3-\phi_1)[q_{k-4}(\mathbf{r}_1, r_{13})-3q_{k-3}(\mathbf{r}_1, r_{13})+3q_{k-2}(\mathbf{r}_1, r_{13})-q_{k-1}(\mathbf{r}_1, r_{13})]\theta(r_{13}-r_{12})\theta(r_{13}-r_{14})
	\notag
	\\
	&\times\theta(r_{13}-r_{15})\theta(r_{13}-r_{16})G_2(1,4)G_2(2,5)G_2(3,6)
	\notag
	\\
	&+1\leftrightarrow 2,
	\label{eq:BBGKY2}
\end{align}
where the notation $1\leftrightarrow 2$ is an abbreviation for all previous terms but with indeces $1$ and $2$ interchanged.
Calculating the time derivative of \cref{eq:G2} we find
\begin{align}
	\partial_t G_2(1,2)=\partial_t P_2(1,2)-P_1(1)\partial_t P_1(2) - P_1(2) \partial_t P_1(1).
	\label{eq:BBGKY3}
\end{align}
Inserting \cref{eq:BBGKY1,eq:BBGKY2} into \cref{eq:BBGKY3} we obtain
\begin{align}
	&\partial_t G_2(1,2)=[-Pe (\cos \phi_1\partial_{x_1}+\sin \phi_1\partial_{y_1})+\frac{1}{2}\partial_{\phi_1}^2]G_2(1,2)
	-\Gamma q_k(\mathbf{r}_1, r_{12})\partial_{\phi_1}K(\phi_2-\phi_1)[P_1(1)P_1(2)+G_2(1,2)]
	\notag
	\\
	&-N\Gamma[q_{k-1}(\mathbf{r}_1, r_{12})-q_{k}(\mathbf{r}_1, r_{12})]P_1(2)\partial_{\phi_1}K(\phi_2-\phi_1)\int d 3 G_2(1,3)\theta(r_{12}-r_{13})
	\notag
	\\
	&-N\Gamma[q_{k-1}(\mathbf{r}_1, r_{12})-q_{k}(\mathbf{r}_1, r_{12})]\partial_{\phi_1}K(\phi_2-\phi_1)P_1(1)\int d 3 G_2(2,3) \theta(r_{12}-r_{13})
	\notag
	\\
	&-N^2\Gamma[q_{k-2}(\mathbf{r}_1, r_{12})-2q_{k-1}(\mathbf{r}_1, r_{12})+q_{k}(\mathbf{r}_1, r_{12})]\partial_{\phi_1}K(\phi_2-\phi_1)\int d 3 d 4 G_2(1,3)G_2(2,4)\theta(r_{12}-r_{13})\theta(r_{12}-r_{14})
	\notag
	\\
	&-N\Gamma \partial_{\phi_1}\int d 3 K(\phi_3-\phi_1)q_{k}(\mathbf{r}_1, r_{13})\theta(r_{12}-r_{13})[G_2(1,2)P_1(3)+P_1(1)G_2(2,3)]
	\notag
	\\
	&-N\Gamma \partial_{\phi_1}\int d 3 K(\phi_3-\phi_1)q_{k-1}(\mathbf{r}_1, r_{13})\theta(r_{13}-r_{12})[G_2(1,2)P_1(3)+P_1(1)G_2(2,3)]
	\notag
	\\
	&-N\Gamma \partial_{\phi_1}\int d 3 K(\phi_3-\phi_1)[q_{k-1}(\mathbf{r}_1, r_{13})-q_{k}(\mathbf{r}_1, r_{13})]\theta(r_{13}-r_{12})[P_1(1)P_1(2)P_1(3)+G_2(1,3)P_1(2)]
	\notag
	\\
	&-N^2\Gamma \partial_{\phi_1}\int d 3 d 4 K(\phi_3-\phi_1)[q_{k-1}(\mathbf{r}_1, r_{13})-q_{k}(\mathbf{r}_1, r_{13})]\theta(r_{12}-r_{13})\theta(r_{13}-r_{14})
	\notag
	\\
	&\times[G_2(1,4)G_2(2,3)+G_2(2,4)P_1(1)P_1(3)+G_2(2,4)G_2(1,3)+G_2(3,4)G_2(1,2)]
	\notag
	\\
	&-N^2\Gamma \partial_{\phi_1}\int d 3 d 4 K(\phi_3-\phi_1)[q_{k-2}(\mathbf{r}_1, r_{13})-q_{k-1}(\mathbf{r}_1, r_{13})]\theta(r_{13}-r_{12})\theta(r_{13}-r_{14})
	\notag
	\\
	&\times[G_2(1,4)G_2(2,3)+G_2(2,4)P_1(1)P_1(3)+G_2(2,4)G_2(1,3)+G_2(3,4)G_2(1,2)]
	\notag
	\\
	&-N^2\Gamma \partial_{\phi_1}\int d 3 d 4 K(\phi_3-\phi_1)[q_{k-2}(\mathbf{r}_1, r_{13})-2q_{k-1}(\mathbf{r}_1, r_{13})+q_{k}(\mathbf{r}_1, r_{13})]\theta(r_{13}-r_{12})\theta(r_{13}-r_{14})
	\notag
	\\
	&\times[G_2(1,4)P_1(2)P_1(3)+G_2(3,4)P_1(1)P_1(2)]
	\notag
	\\
	&-N^3\Gamma \partial_{\phi_1}\int d 3 d 4 d 5 K(\phi_3-\phi_1)[q_{k-2}(\mathbf{r}_1, r_{13})-2q_{k-1}(\mathbf{r}_1, r_{13})+q_{k}(\mathbf{r}_1, r_{13})]\theta(r_{12}-r_{13})\theta(r_{13}-r_{14})\theta(r_{13}-r_{15})
	\notag
	\\
	&\times[G_2(1,4)G_2(2,5)P_1(3)+P_1(1)G_2(2,4)G_2(3,5)]
	\notag
	\\
	&-N^3\Gamma \partial_{\phi_1}\int d 3 d 4 d 5 K(\phi_3-\phi_1)[q_{k-3}(\mathbf{r}_1, r_{13})-2q_{k-2}(\mathbf{r}_1, r_{13})+q_{k-1}(\mathbf{r}_1, r_{13})]\theta(r_{13}-r_{12})\theta(r_{13}-r_{14})\theta(r_{13}-r_{15})
	\notag
	\\
	&\times[G_2(1,4)G_2(2,5)P_1(3)+P_1(1)G_2(2,4)G_2(3,5)]
	\notag
	\\
	&-N^3\Gamma \partial_{\phi_1}\int d 3 d 4 d 5 K(\phi_3-\phi_1)[q_{k-3}(\mathbf{r}_1, r_{13})-3q_{k-2}(\mathbf{r}_1, r_{13})+3q_{k-1}(\mathbf{r}_1, r_{13})-q_{k}(\mathbf{r}_1, r_{13})]\theta(r_{13}-r_{12})\theta(r_{13}-r_{14})\theta(r_{13}-r_{15})
	\notag
	\\
	&\times G_2(1,4)P_1(2)G_2(3,5)
	\notag
	\\
	&-N^4\Gamma \partial_{\phi_1}\int d 3 d 4 d 5 d 6 K(\phi_3-\phi_1)[q_{k-3}(\mathbf{r}_1, r_{13})-3q_{k-2}(\mathbf{r}_1, r_{13})+3q_{k-1}(\mathbf{r}_1, r_{13})-q_{k}(\mathbf{r}_1, r_{13})]\theta(r_{12}-r_{13})\theta(r_{13}-r_{14})
	\notag
	\\
	&\times\theta(r_{13}-r_{15})\theta(r_{13}-r_{16})G_2(1,4)G_2(2,5)G_2(3,6)
	\notag
	\\
	&-N^4\Gamma \partial_{\phi_1}\int d 3 d 4 d 5 d 6 K(\phi_3-\phi_1)[q_{k-4}(\mathbf{r}_1, r_{13})-3q_{k-3}(\mathbf{r}_1, r_{13})+3q_{k-2}(\mathbf{r}_1, r_{13})-q_{k-1}(\mathbf{r}_1, r_{13})]\theta(r_{13}-r_{12})\theta(r_{13}-r_{14})
	\notag
	\\
	&\times\theta(r_{13}-r_{15})\theta(r_{13}-r_{16})G_2(1,4)G_2(2,5)G_2(3,6)
	\notag
	\\
	&+1\leftrightarrow 2.
	\label{eq:BBGKY4}
\end{align}


\subsection{Metric Neighborhoods}

For the metric neighborhood definition, repeated ring-kinetic equations corresponding to \cref{eq:BBGKY1,eq:BBGKY4} have been derived in \cite{KI21}. 
For completeness we write those two equations in our notation
\begin{align}
	\partial_t P_1(\mathbf{r}_1, \phi_1)=&-Pe ( \cos \phi_1 \partial_{x_1}  + \sin \phi_1 \partial_{y_1} ) P_1(1)+ \frac{1}{2}\partial_{\phi_1}^2P_1(1)
	\notag
	\\
	&-N \Gamma \omega_2(\mathbf{r}_1)\int d 2 \theta_{12}\partial_{\phi_1}K(\phi_2-\phi_1)[G_2(1,2)+P_1(1)P_1(2)]
	\notag
	\\
	&-N^2\Gamma [\omega_3(\mathbf{r}_1)-\omega_2(\mathbf{r}_1)]\int d3 d4 \theta_{13}\theta_{14}\partial_{\phi_1} K(\phi_3-\phi_1)[P_1(1)G_2(3,4)+G_2(1,4)P_1(3)]
	\notag
	\\
	&-N^3 \Gamma [\omega_4(\mathbf{r}_1)-2\omega_3(\mathbf{r}_1)+\omega_2(\mathbf{r}_1)]\int d 3 d 4 d 5 \theta_{13}\theta_{14}\theta_{15} \partial_{\phi_1} K(\phi_3-\phi_1)G_2(1,4)G_2(3,5),
	\label{eq:metricBBGKY1}
\end{align}

\begin{align}
	&\partial_t G_2(1,2)=[-Pe (\cos \phi_1\partial_{x_1}+\sin \phi_1\partial_{y_1})+\frac{1}{2}\partial_{\phi_1}^2]G_2(1,2)
	-\Gamma \theta_{12} \omega_2(\mathbf{r}_1) \partial_{\phi_1}K(\phi_2-\phi_1)[P_1(1)P_1(2)+G_2(1,2)]
	\notag
	\\
	&-N \Gamma \theta_{12} [\omega_3(\mathbf{r}_1)-\omega_2(\mathbf{r}_1)]\int d 3 \theta_{13} \partial_{\phi_1} K(\phi_2-\phi_1)[G_2(1,3)P_1(2)+P_1(1)G_2(2,3)]
	\notag
	\\
	&-N^2 \Gamma \theta_{12} [\omega_4(\ro) -2\omega_3(\ro)+\omega_2(\ro)] \int d 3 d 4\theta_{13} \theta_{14} \partial_{\phi_1} K(\phi_2-\phi_1)G_2(1,3)G_2(2,4)
	\notag
	\\
	&-N \Gamma [\theta_{12} \omega_3(\ro) + (1-\theta_{12})\omega_2(\ro)]\int d 3 \theta_{13} \partial_{\phi_1} K(\phi_3-\phi_1) [P_1(1)G_2(2,3)+ G_2(1,2)P_1(3)]
	\notag
	\\
	&-N\Gamma\theta_{12}[\omega_3(\ro)-\omega_2(\ro)] \int d 3 \theta_{13} \partial_{\phi_1} K(\phi_3-\phi_1) [P_1(1)P_1(2)P_1(3)+G_2(1,3)P_1(2)]
	\notag
	\\
	&-N^2\Gamma \big\{\theta_{12}[\omega_4(\ro)-\omega_3(\ro)] + (1-\theta_{12})[\omega_3(\ro)-\omega_2(\ro)]\big\}\int d 3 d 4 \theta_{13} \theta_{14} \partial_{\phi_1} K(\phi_3-\phi_1) 
	\notag
	\\
	&\phantom{-}\times[G_2(1,2)G_2(3,4) + G_2(1,4)G_2(2,3) + G_2(1,3)G_2(2,4) + P_1(1)G_2(2,4)P_1(3)]
	\notag
	\\
	&-N^2\Gamma\theta_{12}[\omega_4(\ro)-2\omega_3(\ro)+\omega_2(\ro)] \int d 3 d 4 \theta_{13} \theta_{14} \partial_{\phi_1} K(\phi_3-\phi_1) [P_1(1)P_1(2)G_2(3,4) + G_2(1,4)P_1(2)P_1(3) ]
	\notag
	\\
	&-N^3\Gamma\big\{\theta_{12}[\omega_5(\ro)-2\omega_4(\ro)+\omega_3(\ro)] + (1-\theta_{12})[\omega_4(\ro)-2\omega_3(\ro)+\omega_2(\ro)] \big\} 
	\notag
	\\
	&\phantom{-}\times\int d 3 d 4 d 5 \theta_{13} \theta_{14} \theta_{15}\partial_{\phi_1} K(\phi_3-\phi_1)[G_2(1,4)G_2(2,5) P_1(3) + P_1(1)G_2(2,4)G_2(3,5) ]
	\notag
	\\
	&-N^3\Gamma\theta_{12}[\omega_5(\ro)-3\omega_4(\ro)+3\omega_3(\ro)-\omega_2(\ro)] \int d 3 d 4 d 5 \theta_{13} \theta_{14} \theta_{15}\partial_{\phi_1} K(\phi_3-\phi_1) G_2(1,4)P_1(2)G_2(3,5) 
	\notag
	\\
	&-N^4\Gamma\big\{\theta_{12}[\omega_6(\ro)-3\omega_5(\ro)+3\omega_4(\ro)-\omega_3(\ro)] + (1-\theta_{12})[\omega_5(\ro)-3\omega_4(\ro)+3\omega_3(\ro)-\omega_2(\ro)]\big\}
	\notag
	\\
	&\phantom{-}\times\int d 3 d 4 d 5 d 6 \theta_{13} \theta_{14} \theta_{15} \theta_{16} \partial_{\phi_1} K(\phi_3-\phi_1) G_2(1,4)G_2(2,5)G_2(3,6)  + 1 \leftrightarrow 2,
	\label{eq:metricBBGKY2}
\end{align}
where $1\leftrightarrow 2$ refers to the repitition of all previous terms with indeces $1$ and $2$ interchanged.
Furthermore we used the notation
\begin{align}
	\theta_{ij}&:=\theta(R-r_{ij})=\begin{cases} 1 \text{ if }R\ge \sqrt{(x_i-x_j)^2+(y_i-y_j)^2} \\ 0 \text{ else},\end{cases}
	\label{eq:thetaij}
	\\
	\omega_k&:=\sum_{s=0}^{\infty}w(k+s)p_{s}(\ro, R).
	\label{eq:omegak}
\end{align}
Thus $\theta_{ij}$ is the indicator function of the condition that particles $i$ and $j$ are neighbors.
The symbols $\omega_k$ denote the expectation value of the weight function $w$ shifted by $k$, see \cref{eq:KSZIresult} for the explicit form of $p_s$.

\section{Clustering Instabilities in the Pair Correlation Function \label{sec:instab}}

Observing the clustering phases $(iii-iv)$ in simulations in Sec. \ref{sec:simulations} we find locally accumulations of particles that move stochastically on large time scales.
Such a phenomenology can not be described in standard mean field theory \footnote{An alternative approach is to consider fluctuating hydrodynamics as a starting point of the theoretical investigations (instead of the microscopic dynamics) and study fluctuations within such a theory, such as e.g. in \cite{PRL126_MCNSTW21,PRL132_SDDT24}.} (setting $G_k\equiv 0$ for $k=2, 3, \dots, N$, see \cref{eq:Gone,eq:Gk} for the definition of $G_k$).
This is because within mean field theory the one particle distribution $P_1$ is always flat in space in the long time limit because the cluster positions are distributed randomly.
Thus, mean field theory can not distinguish the homogeneous phases $(i-ii)$ from the clustering phases $(iii-iv)$.
Evidently, the information of clustering is encoded in higher order correlation functions $G_k$ for $k\ge 2$.

On the other hand, in the apparently homogeneous states $(i-ii)$, mean field, $G_k\approx 0$, seems to be a decent approximation.
Thus, at the onset of clustering there must be an instability that creates nonzero correlation functions $G_k$.
It seems reasonable to assume that the pair correlation function $G_2$ plays a dominant role in driving this instability.
Thus, we are investigating the stability of the $G_2\approx 0$ solution within the repeated ring-kinetic theory derived in Sec. \ref{sec:ringkin}.
Note that even in phases $(i-ii)$ $G_2$ is not exactly zero due to the source terms in \cref{eq:BBGKY4,eq:metricBBGKY2} that depend on the one particle distribution and are present even in the case of a completely disordered $P_1$.
Nevertheless, it is reasonable to assume that correlations are small and consider the linearized (in $G_2$) version of \cref{eq:BBGKY4,eq:metricBBGKY2}
\begin{align}
	\partial_t G_2(1,2) = \mathcal{L}_{ABP}^1 G_{2}(1,2) + \mathcal{L}_{ABP}^2 G_{2}(1,2) + \mathcal{L}_{Coll}[P_1]G_2(1,2) + \mathcal{S}_{Coll}[P_1],
	\label{eq:linearizedGtwo}
\end{align}
where the Active Brownian Particle operator (ABP-operator), acting on each particle individually is 
\begin{align}
	\mathcal{L}_{ABP}^k:=-Pe (\cos \phi_1\partial_{x_k}+\sin \phi_1\partial_{y_k})+\frac{1}{2}\partial_{\phi_k}^2,
	\label{eq:sp_operator}
\end{align}
the linear collision operator $\mathcal{L}_{Coll}[P_1]$ describes the impact of interactions on $G_2$ and $\mathcal{S}_{Coll}[P_1]$ is a source term due to interactions of uncorrelated particles.
Both collision terms, $\mathcal{L}_{Coll}[P_1]$ and $\mathcal{S}_{Coll}[P_1]$ depend on the one particle distribution $P_1$.
We have seen in simulations that the onset of clustering mainly happens when the homogeneous states are polarized, see Sec. \ref{sec:simulations}.
Thus, we assume for the moment that $P_1$ is flat in space and carries orientational order.
However, the linear instabilities of $G_2$ do not depend on the source term $\mathcal{S}_{Coll}$ and $P_1$ resembling some form of inhomogeneities only affect the strength of $\mathcal{L}_{Coll}$ but not the scaling of the dominant instability as we see below.
The detailed form of $\mathcal{L}_{Coll}$ and $\mathcal{S}_{Coll}$ is of no particular importance at the moment, however it is straight forward to derive it by linearizing \cref{eq:BBGKY4,eq:metricBBGKY2} in $G_2$.

\begin{figure}[h]
	\begin{center}
		\includegraphics{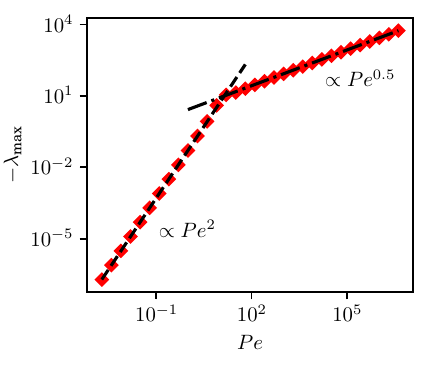}
		\includegraphics{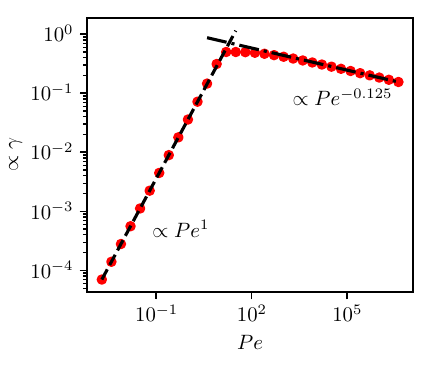}
	\end{center}
	\caption{$(a)$ Decay rate of the slowest decaying eigenmode of the Active Brownian Particle operator (ABP-operator). 
	$(b)$ Excitation of the slowest mode of the ABP-operator due to alignment interactions. 
	The linear system of equations \cref{eq:evp3} was solved numerically for $2\pi k/L_x=2\pi l/L_y=1$ and $j=-195, \dots, 0, \dots 195$ and the solutions is displayed by symbols.
	We find the same asymptotic power laws for other spatial wave vectors. 
	$(b)$ shows the amplitude $|\doublehat{H}_{j=1,kl}|$ of the eigenmode corresponding to the eigenvalue shown in $(a)$. 
	This amplitude is proportional to the excitation of the mode due to alignment interactions. 
	Numerical values of the exponents obtained from fits: $(a)$ small $Pe$-exponent$=2.000000094345308\approx 2$, high $Pe$-exponent$=0.5001363406591083\approx 0.5$, $(b)$ small $Pe$-exponent$=1.000000008775887\approx 1$, high $Pe$-exponent$=-0.12447191974564421\approx-0.125$.
	Dashed and dash-dotted lines represent the power laws obtained from fits for small and high $Pe$, respectively.}
	\label{fig:eigen}
\end{figure}

For the moment we focus on the ABP-operator $\mathcal{L}_{ABP}^k$.
Without any interaction, an active Brownian particle performs a diffusive motion on sufficiently long time scales.
Thus, an initial correlation between two non-interacting ABPs is decaying in time.
Hence, all eigenvalues of the linear operator $\mathcal{L}_{ABP}^1+\mathcal{L}_{ABP}^2$ must be strictly negative.
Because $\mathcal{L}_{ABP}^1$ acts only on the degrees of freedom of particle one and $\mathcal{L}_{ABP}^2$ acts only on the degrees of freedom of particle two, each of the two operators has only strictly negative eigenvalues on its own.
We may write the corresponding eigenvalue problems as
\begin{align}
	\mathcal{L}_{ABP}^1 H_{\lambda}(1)=\lambda H_{\lambda}(1), \qquad \qquad \mathcal{L}_{ABP}^2 J_{\mu}(2)= \mu J_{\mu}(2).
	\label{eq:evp1}
\end{align}
In order to solve it, we consider the spatial Fourier transform of the function $H$
\begin{align}
	H(x_1, y_1, \phi_1)=\sum_{k,l} \hat{H}_{kl}(\phi_1) \exp[i(2\pi/L_x k x_1+2\pi/L_y l y_1)].
	\label{eq:spatial_fourier_H}
\end{align}
Inserting the Fourier transform \cref{eq:spatial_fourier_H} into the eigenvalue problem \cref{eq:evp1} we observe that there is no interaction between different spatial modes.
Thus, we can consider a separate problem for each mode
\begin{align}
	[iPe 2 \pi(k/L_x \cos \phi_1 +l/L_y\sin \phi_1)+ 1/2 \partial_{\phi_1}^2]\hat{H}_{kl,\lambda_{kl}}(\phi_1)=\lambda_{kl}\hat{H}_{kl,\lambda_{kl}}(\phi_1).
	\label{eq:evp2}
\end{align}
Thus, the eigenvalue problem does not depend on $Pe$, $k/L_x$ and $l/L_y$ separately but only on the combinations $Pe k/L_x$ and $Pe l/L_y$.
We employ the Fourier transform of the orientational degree of freedom
\begin{align}
	\hat{H}_{kl}(\phi_1)=\sum_j \doublehat{H}_{jkl}\exp(i2 \pi j \phi_1)
	\label{eq:fourier_H}
\end{align}
to rewrite the eigenvalue problem \cref{eq:evp2} once more as
\begin{align}
	\lambda_{kl}\doublehat{H}_{jkl,\lambda_{kl}}=- 1/2 (2\pi)^2 j^2\doublehat{H}_{jkl,\lambda_{kl}} + Pe \pi(ik/L_x +i l/L_y)\doublehat{H}_{j-1,k,l,\lambda_{kl}}+ Pe \pi(i k/L_x - l/L_y)\doublehat{H}_{j+1,k,l,\lambda_{kl}}.
	\label{eq:evp3}
\end{align}
For fixed spatial wave vectors, $k$, $l$, \cref{eq:evp3} displays a system of linear equations that we solve numerically in order to obtain eigenvalues and eigenvectors.
In Fig. \ref{fig:eigen} $(a)$ we show the largest eigenvalue as a function of \Peclet number.
For large \Peclet numbers we find that it is proportional to the square root of the \Peclet number,
\begin{align}
	-\lambda_{\text{max}}\propto Pe^{0.5}
	\label{eq:ev_high_pe}
\end{align}
and for small \Peclet numbers it is proportional to the square of the \Peclet number, 
\begin{align}
	-\lambda_{\text{max}}\propto Pe^{^2}.
	\label{eq:ev_small_pe}
\end{align}
The data of Fig. \ref{fig:eigen} correspond to one particular choice of the spatial wave vectors, however, we do find the same power laws for all choices of these spatial wave vectors.

In general, we may write the pair correlation function in terms of the eigenfunctions as
\begin{align}
	G_{2}(1,2) = \,\, \mathclap{\displaystyle\int}\mathclap{\textstyle\sum} \,\,\,\, d \lambda d \mu f(\lambda, \mu) H_{\lambda}(1) J_{\mu}(2),
	\label{eq:decomp_gtwo}
\end{align}
where the symbol $\,\,\,\mathclap{\displaystyle\int}\mathclap{\textstyle\sum} \,\,\,\, d \lambda d \mu $ denotes an integral in the case of an infinite domain and a sum in the case of a final domain.

Considering the limit of high \Peclet numbers the decay of correlations due to the ABP-operator is strong and hence the slowest decaying eigenmode is dominating
\begin{align}
	G_{2, \text{max}}=H_{\lambda_{\text{max}}}(1)J_{\mu_{\text{max}}}(2),
	\label{eq:dominant_mode}
\end{align}
where in fact $\lambda_{\text{max}}=\mu_{\text{max}}$.
We define the projection operators $\mathcal{P}_{1,\lambda}$ and $\mathcal{P}_{2,\mu}$ that project the dependence of $G_2(1,2)$ on $1$ onto $H_{\lambda}(1)$ and the dependence on $2$ onto $J_{\mu}(2)$. 
Assuming that the eigenvalue problem of the linear operator $\mathcal{L}_{ABP}^1+\mathcal{L}_{ABP}^2+\mathcal{L}_{Coll}[P_1]$ is dominated by the eigenstates of the ABP-operator, we find approximately for the largest eigenvector
\begin{align}
	(\mathcal{L}_{ABP}^1+\mathcal{L}_{ABP}^2+\mathcal{L}_{Coll}[P_1]) G_{2, \text{max}}\approx (\lambda_{\text{max}}+\mu_{\text{max}}+\gamma\Gamma)G_{2, \text{max}} = (2\lambda_{\text{max}}+\gamma\Gamma)G_{2, \text{max}},
	\label{eq:evp4}
\end{align}
with $\Gamma$ being the interaction strength incorporated into $\mathcal{L}_{Coll}$ and $\gamma$ is given by
\begin{align}
	\gamma \Gamma G_{2,\text{max}}=  \mathcal{P}_{1, \lambda_{\text{max}}} \mathcal{P}_{2, \mu_{\text{max}}} \mathcal{L}_{Coll}[P_1] G_{2,\text{max}}.
	\label{eq:evp5}
\end{align}
We might estimate the dependence of $\gamma$ on $Pe$ by thinking about the action of $\mathcal{L}_{Coll}[P_1]$ physically.
In general the collision operator creates polar alignment between the colliding particles.
There are terms that act on $\phi_1$ and $\phi_2$ and thus create alignment between those two particles.
Furthermore, there are terms that act on $\phi_{1/2}$ and $\phi_3$, where $3$ is an uncorrelated particle, polarized according to $P_1$, see \cref{eq:BBGKY4,eq:metricBBGKY2}.
Thus, the orientational Fourier Transform of the distribution of $\mathcal{L}_{Coll} G_{2, \text{max}}$ has its major contributions for angular mode numbers $j_{1/2}=\pm 1$.
Projecting onto $\mathcal{P}_{1,\lambda_{\text{max}}}$ and $\mathcal{P}_{2,\mu_{\text{max}}}$ gives then prefactors proportional to the amplitudes $|\doublehat{H}_{j=\pm 1}|$ or $|\doublehat{J}_{j=\pm 1}|$.
Because these amplitudes are small, the dominant contribution to $\gamma$ comes from interaction terms where alignment acts only on one of the coordinates $\phi_{1/2}$ such that the corresponding prefactor due to the projection operators is proportional to $|\doublehat{H}_{j=\pm 1}|$ or $|\doublehat{J}_{j=\pm 1}|$ but not to $|\doublehat{H}_{j=\pm 1}||\doublehat{J}_{j=\pm 1}|$.
Hence we have
\begin{align}
	\gamma \propto |\doublehat{H}_{j=1}|.
	\label{eq:ev_collision}
\end{align}
In Fig. \ref{fig:eigen} $(b)$ we display the amplitude of the mode $\doublehat{H}_{j=1}$ as a function of \Peclet number $Pe$.
We find the power law behavior
\begin{align}
	|\doublehat{H}_{j=1}|\propto \gamma \propto Pe^{-0.125}
	\label{eq:interaction_high_pe}
\end{align}
for high \Peclet numbers and
\begin{align}
	|\doublehat{H}_{j=1}|\propto \gamma \propto Pe^{2}
	\label{eq:interaction_small_pe}
\end{align}
for small \Peclet numbers.
According to \cref{eq:evp4} the linear instability sets on above $2\lambda_{\text{max}}+\Gamma\gamma=0$.
Combining the high $Pe$ behavior of $\lambda_{\text{max}}$, \cref{eq:ev_high_pe}, and of $\gamma$, \cref{eq:interaction_high_pe}, we obtain for the onset of the instability
\begin{align}
	Pe/\Gamma^{1.6}=const.
	\label{eq:instability_high_pe}
\end{align}

Note that in parts of the above discussion, we considered the eigenstates $H_{\lambda}$ of the ABP-operator in Euclidean space, whereas in other parts, for technical reasons, we considered it in (spatial) Fourier space.
In fact, the relevant eigenstate might be a linear superposition of several spatial Fourier modes.
However, because all spatial Fourier modes follow the same power law for large and for small \Peclet numbers, the scaling law of the instability, \cref{eq:instability_high_pe} is nevertheless valid.

In general, the linear instability does not provide detailed information on the resulting pattern, that is determined by nonlinear effects.
In fact, there is a feedback of large pair correlations $G_2$ onto the one particle distribution $P_1$ that determines global polarization.
In the simulations of Sec. \ref{sec:simulations} we find polarization in some cases under the presence of clustering and in other cases not.
Nonlinear analysis is needed to understand this behavior, which goes beyond the scope of this work.

It should be noted that in the above discussion, the notion of high \Peclet numbers has to be interpreted in a wide sense, because the actual large parameter is the product of \Peclet number and wave vector, see \cref{eq:evp3}.
The largest possible wave number roughly corresponds to the wave length matching the length scale of the collision integrals, because for much smaller length scales (highly oscillating over the interaction range) no correlations are created due to collisions.
Furthermore, we observe from Fig. \ref{fig:eigen}, that the large $Pe$ behavior is already present at not too large values of $Pe$, there however, it does not perfectly match the power law.
In fact, we do find the clustering instability in simulations also at small \Peclet numbers, see Sec. \ref{sec:simulations}, but in this case, the onset of the stability does not perfectly follow the power law \cref{eq:instability_high_pe}.

Considering the limit of small \Peclet numbers, or small wave vectors, there is another type of instability, see Fig. \ref{fig:eigen}.
The onset of this instability scales, according to \cref{eq:ev_small_pe,eq:interaction_small_pe} as
\begin{align}
	Pe/\Gamma=const.
	\label{eq:instability_small_pe}
\end{align}
Indeed we find large scale patterns, phase $(v)$, with the corresponding scaling in simulations, see Sec. \ref{sec:simulations}.

\section{Discussion\label{sec:discussion}}

We study active Brownian particles (ABPs) in two dimensions that are subject to alignment interactions with neighboring particles.
Neighborhoods are either defined via an interaction distance (metric interactions) or by specifying the number of neighbors, considering the closest surrounding particles (topological interactions).
In both cases we observe a clustering instability that leads to the formation of small groups of particles (clusters) that move coherently.
In case of topological interactions, the clustering was observed before \cite{ZHR22}, however, our study shows that the clustering instability is a general feature of interacting ABPs that is driven by the eigenstates of the ABP-operator. 
Its occurrence does not depend on the details of the interaction.

In general, the behavior of interacting ABPs is characterized by two essential mechanisms: self-propulsion and interaction.
The strength of self-propulsion, or activity, is described by the \Peclet number $Pe$, whereas the interaction can be quantified by a coupling strength $\Gamma$.
We show by means of repeated ring-kinetic theory that the clustering instability is related to an instability of the pair correlation function.
In fact, the clustering state can not be understood by standard mean field theory because the clusters move stochastically and thus on average all one-particle-fields are distributed homogeneously in steady state.
Thus, the consideration of correlations is necessary in order to describe the clustering state.
We show within kinetic theory, that the onset of clustering depends only on the combination of activity and coupling of the form $Pe/\Gamma^{1.6}$ within the high activity limit.
This prediction is confirmed by agent-based simulations for both, metric and topological interactions.
It should be mentioned that for topological interactions \cite{ZHR22} came to the conclusion that the onset of clustering depends on $Pe/\Gamma^{\alpha}$ with $\alpha\approx 1.5$.
This exponent was extracted purely from simulation data.
There are several possible explanations for this small discrepancy.
First, \cite{ZHR22} considered slightly different interactions and simulations have been performed for smaller systems ($N=2500$).
Second, the exponent was extracted from data for small, intermediate and large \Peclet numbers.
We derive the exponent $\alpha=1.6$ for high \Peclet numbers.
For smaller \Peclet numbers the clustering instability exists as well but it does not strictly follow the power law for large $Pe$.

In the limit of small \Peclet numbers, we find at high coupling a different, long-wavelength instability that dominates over the clustering instability within kinetic theory.
The presence of this stability is confirmed in agent-based simulations and manifests itself through the creation of large scale polarized patterns.
Due to the long wavelength of the instability, local correlations on the length scale of the interaction distance vanish within this pattern.
The onset of this instability depends on the ratio of parameters $Pe/\Gamma$.
This relation is confirmed in simulations for metric and topological interactions.

From a technical point of view, the derivation of repeated ring-kinetic theory is significantly more challenging then mean field theory that corresponds to the study of solution of the nonlinear Fokker-Planck-equation.
In contrast, here we consider the first two members of the BBGKY-hierarchy \cite{\bbgky} and only approximate correlations between three or more particles as negligible.
In analogy to the nonlinear Fokker-Planck-equation, here, we obtain a set of two coupled nonlinear equations for the one-particle distribution and for the pair correlation function.
The repeated ring-kinetic theory for metric interactions has been derived in a previous study \cite{KI21}.
Here, we complement this work by giving the full derivation for topological interactions.
Although the derivation is elementary, some specifically tailored combinatorial concepts have to be developed and preliminary works on the spatial distribution of correlated particles are employed \cite{\correlations}.
The repeated ring-kinetic theory developed here, might be of future use for quantitative studies on the topological model, similar to the previous work \cite{KI21} on the metric model.

Here, we study the eigenstates of the ABP-operator semi-numerically in order to extract the relevant power laws of the eigenvalues for large and small \Peclet numbers.
Possibly, this eigenvalue problem could be solved fully analytically at least in the limits of small or large self-propulsion.
In that way further insights on the clustering instability might be gained.

In summary, we report a clustering mechanism for interacting active Brownian particles based on an instability of the pair correlation function.
The mechanism mainly depends on the eigenstates of the active Brownian particle operator and the details of the interactions are of no importance.
Thus, the mechanism is of relevance to a large class of active matter, interacting self-propelled particles.

\acknowledgements{The author thanks T. Ihle for bringing the work \cite{ZHR22} to his attention and thus motivating this study.
The author thanks Universitätsrechenzentrum Greifswald for providing computational resources.
The author acknowledges support of this work by a 'María Zambrano'-grant at Universitat de Barcelona financed by the Spanish Ministry of Science and European Union (EU next generation).}

\end{document}